\documentclass{aa}

\usepackage{graphicx}
\usepackage{txfonts}
\usepackage[bookmarks=true,
            bookmarksopen=true, 
            pdfstartview = FitV,
            pdfpagelayout = SinglePage,
            colorlinks = true,
            linkcolor=black,
            urlcolor = blue,
            pdfborder = {0 0 0}]{hyperref}

\usepackage[dvipsnames]{xcolor}
\usepackage{color}
\begin{document}

   \title{Revealing the burning and soft heart of the bright bare AGN ESO\,141-G55: X-ray broadband and SED analysis}
 \titlerunning{Revealing the burning and soft heart of the bright bare AGN ESO\,141-G55}
            \authorrunning{Porquet et al.}
   \author{D.\ Porquet \inst{1}
         \and
         J.~N.\ Reeves\inst{2,3}
          \and
           S.\ Hagen\inst{4}
           \and            
           A.\ Lobban\inst{5}
          \and
          V.\ Braito\inst{3,2}
          \and
          N.\ Grosso\inst{1}
            \and
           F.\ Marin\inst{6}
         }
   \institute{Aix Marseille Univ, CNRS, CNES, LAM, Marseille, France 
              \email{delphine.porquet@lam.fr}
        \and  Department of Physics, Institute for Astrophysics and Computational
Sciences, The Catholic University of America, Washington, DC 20064, USA
         \and INAF, Osservatorio Astronomico di Brera, Via Bianchi 46 I-23807 Merate
 (LC), Italy
         \and
          Centre for Extragalactic Astronomy, Department of Physics, University of Durham, South Road, Durham DH1 3LE, UK
          \and European Space Agency (ESA), European Space Astronomy Centre (ESAC), E-28691 Villanueva de la Cañada, Madrid, Spain 
           \and
            Universit{\'e} de Strasbourg, CNRS, Observatoire Astronomique de Strasbourg, UMR 7550, 67000 Strasbourg, France  
 }
   \date{Received , 2024; accepted , 2024}

   \abstract
  {ESO 141-G55 is a nearby X-ray bright broad-line Seyfert\,1 (BLS1), which has been classified as a bare active galactic nucleus (AGN) due to the lack of warm absorption along its line-of-sight, providing an unhampered view into its disc-corona system.}
     {We aim to probe its disc-corona system thanks to the first simultaneous {\sl XMM-Newton} and {\sl NuSTAR} observation obtained on October 1--2, 2022.}
   {We carry out the X-ray broadband spectral analysis to determine the dominant process(es) at work, as well as the spectral energy distribution (SED) analysis  to determine the disc-corona properties.}
   {The simultaneous broadband X-ray spectrum of ESO\,141-G55 is characterised by the presence of a prominent smooth soft X-ray excess, a broad Fe\,K$\alpha$ emission line and a significant Compton hump. The high-resolution RGS spectra confirmed the lack of intrinsic warm-absorbing gas along our line of sight in the AGN rest frame, confirming that it is still in a bare state. However, soft X-ray emission lines are observed indicating substantial warm gas out of our line of sight. 
   The intermediate inclination of the disc-corona system, $\sim$43$^{\circ}$, may offer us a favourable configuration to observe  ultra-fast outflows from the disc, but none is found in this 2022 observation, contrary to a previous 2007 {\sl XMM-Newton} one. Relativistic reflection alone on a standard disc is ruled out from the X-ray broadband analysis, while a combination of soft and hard Comptonisation by a warm and hot corona ({\sc relagn}), plus relativistic reflection ({\sc reflkerrd}) reproduces its SED quite well. 
  The hot corona temperature is very hot, $\sim$140\,keV, much higher than about 80\% of the AGNs, whereas the warm corona temperature, $\sim$0.3\,keV, is similar to the values found in other sub-Eddington AGNs. ESO\,141-G55 is accreting at a moderate  Eddington accretion rate ($\sim$10--20\%).} 
{Our analysis points to a significant contribution of an optically-thick warm corona to both the soft X-ray and UV emission in ESO\,141-G55, adding to growing evidence that AGN accretion (even at moderate accretion rate) appears to  deviate from standard disc theory.}
   \maketitle

%
%________________________________________________________________

\section{Introduction}\label{sec:Introduction}

In the so-called ‘bare’ AGN, our line of sight intercepts very little or no warm X-ray absorption. Thus, bare AGN allow a clean, unimpeded view into the innermost core of the nucleus. As a result, both the continuum (soft X-ray excess and Compton hump) and the emission lines can be measured safely, preventing the complexities introduced from modelling complex absorption, providing then a robust probe of the disc-corona system.

Importantly, broadband X-ray spectra are mandatory to determine the physical processes that produce the soft X-ray excess: relativistic reflection resulting from the illumination of the accretion disc by a hot corona,
or Comptonisation of seed photons from the accretion disc by a warm and a hot corona, or a combination thereof \citep[e.g.,][]{Magdziarz98,Porquet04a,Crummy06,Bianchi09,Fabian12,Done12,Gliozzi13,Porquet18,Petrucci18,Porquet19,Gliozzi20,Waddell20,Porquet21,Porquet24}. So far, only a handful of very bright bare AGN have been observed simultaneously by {\sl NuSTAR}  and {\sl XMM-Newton} with a very high signal-to-noise ratio (\object{Fairall\,9}, \object{Ark \,120}, \object{Mrk 110}, and {\,TON\,S180}) allowing us to demonstrate that relativistic reflection alone onto a standard disc cannot reproduce their X-ray broadband spectra. Instead, a combination of soft and hard Comptonisation (by a warm and hot corona) and relativistic reflection reproduces their SEDs quite well \citep{Matt14,Porquet18,Lohfink16,Matzeu20,Porquet21}.

\begin{table*}[t!]
\caption{Observation log of the {\sl XMM-Newton} and {\sl NuSTAR} dataset for ESO\,141-G55.
}
\begin{tabular}{c@{\hspace{15pt}}l@{\hspace{15pt}}l@{\hspace{15pt}}l@{\hspace{15pt}}c@{\hspace{15pt}}c@{\hspace{15pt}}c}
\hline \hline
\multicolumn{1}{c}{Mission} & \multicolumn{1}{c}{Obs.\,ID} & \multicolumn{1}{l}{Obs.\ date}  & \multicolumn{1}{l}{Exp.$^{(a)}$}& \multicolumn{1}{c}{\sl XMM-Newton} & mean CR$^{(b)}$ & labelled in the text \\
        &          &   \multicolumn{1}{l}{(yyyy-mm-dd)}       & \multicolumn{1}{l}{(ks)}      &  \multicolumn{1}{c}{window mode} &        count\,s$^{-1}$                \\
\hline
%\hline
\multicolumn{6}{c}{This work}\\
\hline
{\sl NuSTAR}     & 60801011002 & 2022-10-01  & 124.1 &  & 0.74 (FPMA) & 2022\\
                 &             &                     & 122.3 &    & 0.69 (FPMB) & 2022\\
{\sl XMM-Newton} & 0913190101 & 2022-10-01 & 122.6 (72.4) & Small Window & 24.1 (pn) & 2022\\
\hline
\multicolumn{6}{c}{Previous observations}\\
\hline
{\sl NuSTAR}     & 60201042002 & 2016-07-15   & 93.0 &  &  0.61 (FPMA)& 2016\\
                 &             &                       &        92.9  &   &  0.56 (FMPB) &2016\\
{\sl XMM-Newton} & 0503750101 & 2007-10-30  & 77.7 (67.7) & Full Frame$^{(c)}$ & 6.3$^{(d)}$ (pn)& 2007\#4\\
{\sl XMM-Newton} & 0503750501 & 2007-10-12  & 22.3 (16.6) & Full Frame$^{(c)}$ &  5.6$^{d)}$ (pn)\\
{\sl XMM-Newton} & 0503750401 & 2007-10-11  & 21.0 (18.3) &  Full Frame$^{(d)}$ & 6.2$^{(d)}$ (pn)\\
{\sl XMM-Newton} & 0503750301 & 2007-10-09 & 20.0 (17.5) &  Full Frame$^{(c)}$ & 6.8$^{(d)}$ (pn)\\
{\sl XMM-Newton} & 0101040501 & 2001-10-09 & 53.8  &  Full Frame$^{(c)}$ & n.a.$^{(e)}$\\
\hline
\end{tabular}
\label{tab:log}
\flushleft
\small{\textit{Notes.} 
$^{(a)}$ Total exposure and in parenthesis the net pn exposure time after correction of background flares and deadtime. 
$^{(b)}$ Mean source count rate over 0.3--10 keV for {\sl XMM-Newton} and over 3--79\,keV for {\sl NuSTAR}. 
$^{(c)}$ These datasets suffer for very strong pile-up due to the use of the Full Frame mode. 
$^{(d)}$ Net count rate inferred using an annular extraction region to minimise the very strong pile-up (see text).
$^{(e)}$ Only MOS data are available and are thus not used in the present work. 
}
\end{table*}

Here, our objective is to expand this important study to the broad-line Seyfert\,1 (BLS1) bare AGN \object{ESO\,141-G55} ($z$=0.0371), thanks to the first simultaneous {\sl XMM-Newton} and {\sl NuSTAR} observation performed in October 1--2, 2022 (Table~\ref{tab:log}). Its supermassive black hole mass (log\,(M/M$_{\odot}$)=8.1$\pm$0.1) and the inclination angle of its accretion disc (which ranges from 33$^{\circ}$ to 42$^{\circ}$ for a maximal and a non-spinning black hole, respectively) have been inferred from accretion disc fits to the UV continuum and the H$\beta$ emission line \citep{Rokaki99}.
ESO\,141-G55 shows a high bolometric luminosity of L$_{\rm bol}$$\sim$3$\times$10$^{45}$\,erg\,s$^{-1}$ (L$_{\rm 2-10\,keV}$$\sim$10$^{44}$\,erg\,s$^{-1}$) thus corresponding to an Eddington accretion rate of 10--20\%.
It is one of the brightest hard X-ray AGN seen by {\sl Swift}-BAT
(F$_{\rm 14-195 keV}$$\sim$6$\times$10$^{-11}$\,erg\,cm$^{-2}$\,s$^{-1}$), 
with so far a very moderate X-ray flux variability of $\sim$20\% reported over 4--5 decades with: {\sl Ariel-V} \citep{Cooke78,Elvis78}, {\sl ASCA} \citep{Reynolds97}, {\sl XMM-Newton} \citep{Gondoin03},  {\sl Swift} \citep{Oh18}, and {\sl NuSTAR} \citep{Ezhikode20,Panagiotou20,Ghosh20,Akylas21,Kang22}. 
The hard X-ray spectrum of ESO\,141-G55 exhibits a very high-energy cutoff of a few hundred keV, as measured from a {\sl NuSTAR} observation performed in 2016 (Table~\ref{tab:log}), showing the presence of a very hot corona \citep[e.g.,][]{Ezhikode20,Panagiotou20,Akylas21,Kang22}.  \\

ESO\,141-G55 has previously been observed five times by {\it XMM-Newton} with $F_{\rm 0.3-10\,keV}$=5.0--6.3$\times$10$^{-11}$\,erg\,cm$^{-2}$\,s$^{-1}$, once in 2001 and four times in 2007 (Table~\ref{tab:log}), but for all of them the full frame (FF) science mode was applied to the EPIC (MOS and pn) cameras leading to very strong pile-up.
Interestingly, in all these observations, ESO\,141-G55 exhibits a smooth prominent soft X-ray excess, with no signature from warm absorbing gas in its rest-frame from the spectral analysis of the reflection grating spectrometer (RGS) data, classifying it as a bare AGN \citep{Gondoin03,Boissay16,Ghosh20}. 
The only reported X-ray absorption from warm gas towards ESO\,141-G55 arises from foreground \ion{N}{vi} and \ion{O}{vii} associated with our Galaxy with modest ionic columns \citep[N(\ion{N}{vi})$\sim$4$\times$10$^{17}$\,cm$^{-2}$ and N(\ion{O}{vii})$\sim$10$^{16}$\,cm$^{-2}$;][]{Fang15,Gatuzz21}.

The intermediate inclination of its disc-corona system may offer us a favourable configuration to intercept the blueshifted absorption signature(s) of a possible equatorial disc wind (ultra-fast outflows, UFO) in the Fe\,K energy band ($\sim$6-7\,keV), as reported in at least about 40\% of the local AGN \citep[e.g.][]{Tombesi10,Gofford13}. Indeed, during the 2001 {\sl XMM-Newton} observation (see Table~\ref{tab:log}), an apparent absorption feature at 7.6$\pm$0.1\,keV interpreted as an Fe\,K$\alpha$ edge ($\tau$=0.3$\pm$0.1) was observed with the MOS cameras. Unfortunately, no {\sl XMM-Newton}-pn data were acquired, preventing any confirmation of this possible absorption feature with a better sensitivity above 6\,keV. Moreover, the MOS data were obtained using the full frame window leading to heavy pile-up. Applying an annulus extraction region (from 11$\arcsec$ and 86$\arcsec$) to minimise the pile-up effect significantly decreases the signal-to-noise ratio of the MOS spectrum \citep{Gondoin03}. However, during the {\sl XMM-Newton}-pn observation on 30 October 2007 (that is, the longest observation in 2007, Table~\ref{tab:log}, hereafter labelled 2007\#4), after minimising the strong pile-up effect, a highly blue-shifted absorption feature was reported at $\sim$8.4\,keV with $v_{\rm out}$ =-0.18$\pm$0.01\,$c$ at a $\geq$4$\sigma$ confidence level \citep{DeMarco09}. No absorption line was observed in the three other shorter {\sl XMM-Newton} observations, which were only performed about 3 weeks earlier. \\

In this paper, our aim is to probe the disc-corona system of ESO\,141-G55 and its close vicinity,  using its first simultaneous {\sl XMM-Newton} and {\sl NuSTAR} observation performed on October 1--2, 2022. 
 We also perform, when relevant, some comparisons with the previous reprocessed 2007 {\sl XMM-Newton} and 2016 {\sl NuSTAR} observations. 
In Section~\ref{sec:obs}, the data reduction and analysis methods of the datasets are presented. 
The X-ray spectral analysis is reported in Section~\ref{sec:mainX}, using the high-resolution {\sl XMM-Newton}-RGS data, as well as the {\sl XMM-Newton}-pn and NuSTAR data. 
The SED analysis (UV to hard X-rays) is reported in Section~\ref{sec:SED2022}. 
Finally, the main results are summarised and discussed in Section~\ref{sec:discussion}.

\section{Observations, data reduction and analysis method}\label{sec:obs}

The log of this first simultaneous {\sl XMM-Newton} and {\sl NuSTAR} observation of ESO\,141-G55
(NuSTAR cycle-8; PI: D.\ Porquet) used in this work is reported in Table~\ref{tab:log}.
Previous observations with {\sl XMM-Newton} and {\sl NuSTAR} are also listed. 

\subsection{{\sl XMM-Newton} data reduction}\label{sec:xmm}

{\sl XMM-Newton} data were reprocessed with the Science Analysis System (SAS, version 20.0.0), applying the
latest calibration available on April 25, 2023. 
Only the EPIC-pn \citep{Struder01} data are used (selecting the event patterns 0--4, that is to say, single and double events) since they have a much better sensitivity above $\sim$6\,keV than MOS data. Therefore, we did not use the 2001 observation. 
The 2022 EPIC pn observation was adequately operated in the Small Window mode and then the pn data are not subject to pile-up.
The 2022 pn spectrum was extracted from a circular region centered on ESO\,141-G55, with a radius of 35${\arcsec}$ to avoid the edge of the chip. 
The four 2007 pn spectra, in order to minimise the heavy pile-up due to the use of the Full Frame window mode, were extracted from an annulus region centred on the source with an outer radius 35${\arcsec}$ and with an inner radius of 11$\arcsec$ (ObsID: 0503750401), 12$\arcsec$ (ObsID: 0503750101) and 13$\arcsec$ (ObsIDs: 0503750301 and 0503750501). An annulus extraction region results in pn count rates about four times lower in 2007 compared to 2022 (Table~\ref{tab:log}) and, thus, in a significant decrease of the signal-to-noise ratio for all 2007 pn spectra. We note that neglecting such an important pile-up effect would considerably impact the spectral analysis, for example producing an artificial `harder when brighter' X-ray behavior of the source. 
All background spectra were extracted from a rectangular region that contains no (or negligible) source photons. 
 Total and net exposure times of the pn (obtained after correction for dead time and background flaring) are reported in Table~\ref{tab:log}. 
All redistribution matrix files (rmf) and ancillary response files (arf) were generated with the SAS tasks {\sc rmfgen} and {\sc arfgen}, and were binned in order to over-sample the instrumental resolution by at least a factor of four, with no impact on the fit results. For the arf calculation, the recent option {\sc applyabsfluxcorr}=yes was  applied, which allows for a correction of the order of 6–8\% between 3 and 12\,keV in order to reduce differences in the spectral shape between {\sl XMM-Newton}-pn and {\sl NuSTAR} spectra (F.\ Fürst 2022, XMM-CCF-REL-388, XMM-SOC-CAL-TN-0230)\footnote{https://xmmweb.esac.esa.int/docs/documents/CAL-SRN-0388-1-4.pdf}. Finally, the background-corrected pn spectra were binned to have a signal-to-noise ratio greater than four in each spectral channel.\\
 
%-----------------------------Figure --------------------------------
\begin{figure}[t!]
\begin{tabular}{c}
\includegraphics[width=0.9\columnwidth,angle=0]{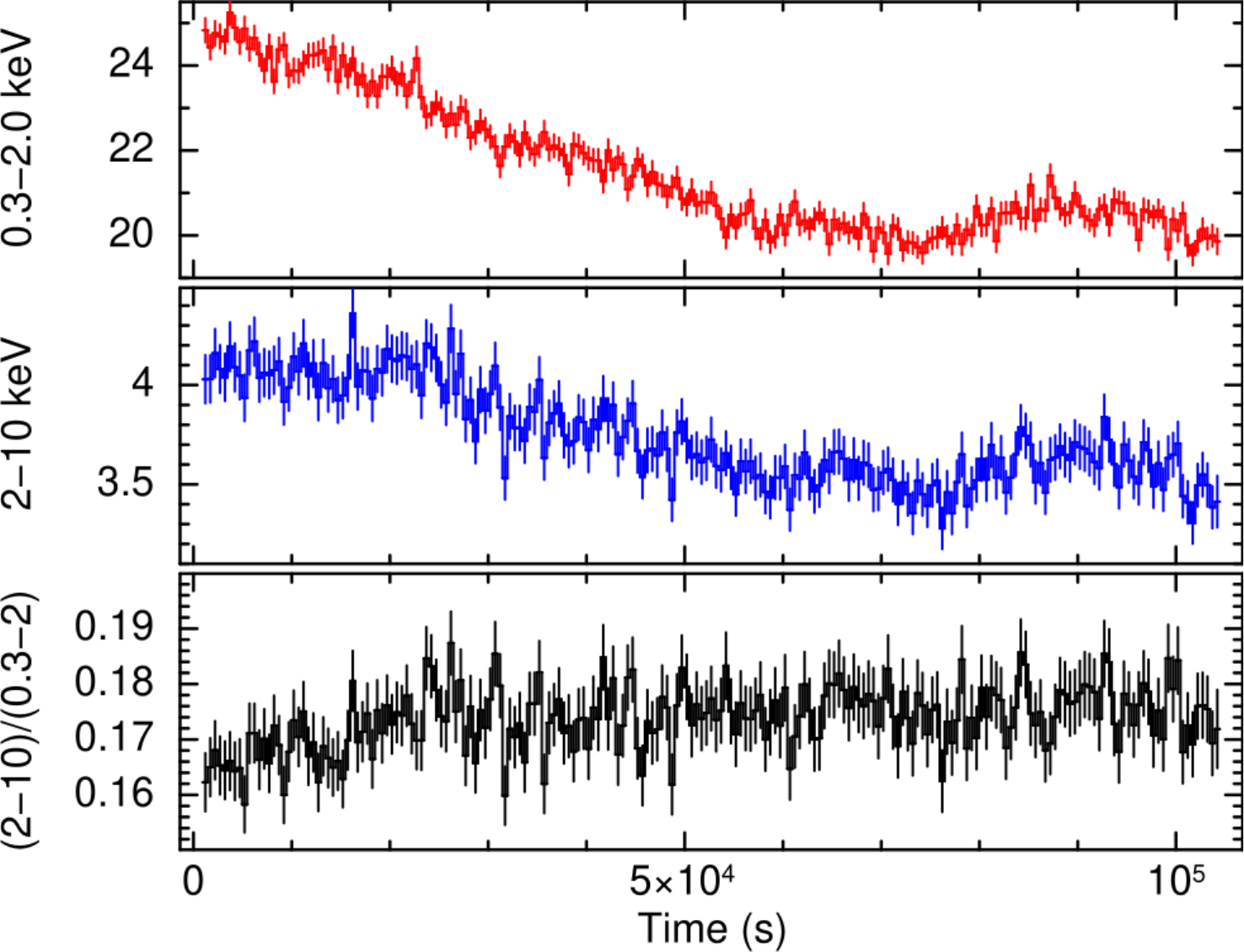}\\
\end{tabular}
	\caption{The 2022 {\sl XMM-Newton}-pn light curves of ESO\,141--G55, with time bins of 500\,s, in the 0.3--2\,keV (top panel) and 2-10\,keV (bottom panel) energy ranges and the corresponding hardness ratio, 2--10\,keV/0.3--2\,keV.}
 \label{fig:pnlc}
\end{figure}
%-------

Figure~\ref{fig:pnlc} displays the 0.3--2\,keV (top panel) and 2--10\,keV (middle panel) pn count rate curves during the 2022 campaign, built using the {\sl epiclccorr} {\sc SAS} tool.  A smooth decrease, most prominent in the soft X-ray band, of the source flux is observed during the first half of the observation. However, there is only very little variation in the hard/soft ratio (bottom panel) mainly in the first 20\,ks, where the AGN displays a slight `softer when brighter' behaviour. Performing a simplified modeling of the pn spectra during the `high' and `low' fluxes, we found that the variability is solely driven by the flux change, since no significant spectral shape variability is observed. 
The light curves for the four 2007 {\sl XMM-Newton}-pn observations are reported in Fig.~\ref{fig:2007lc}. Even if in the first and third observations notable count rate variability is observed the hardness ratio indicates no significant spectral variability during each observation.  In this work, we used the time-averaged pn spectra for each epoch.\\

 RGS spectra of ESO\,141$-$G55, for the 2022 and the 2007\#4 observations were reprocessed and analysed. The individual RGS\,1 and RGS\,2 spectra at each epoch were combined into a single merged spectrum using the \textsc{SAS} task \textsc{rgscombine}. This resulted in a total count rate from the combined RGS spectrum at each epoch of 1.288$\pm$0.003\,cts\,s$^{-1}$ over an exposure time of 109.4\,ks for 2022, and of 1.600$\pm$0.005\,cts\,s$^{-1}$ over an exposure time of 79.3\,ks for 2007\#4. The spectra were binned to sample the RGS spectral resolution \citep{denHerder01}, by adopting a constant wavelength binning of $\Delta\lambda$=0.05\,\AA\ per spectral bin over the wavelength range from 6--36\,\AA. This binning also ensures a minimum signal-to-noise ratio per bin of 10 over the RGS energy range and thus $\chi^2$ minimisation was used for the spectral fitting procedure. \\
 
 UV data from the XMM-Newton Optical-UV Monitor (hereafter OM; \citealt{Mason01}) were processed for the 2022 and 2007\#4 observations using the SAS script {\sc omichain}. This script takes into account all necessary calibration processes (e.g. flat-fielding) and runs a source detection algorithm before performing aperture photometry (using an extraction radius of 5.7$\arcsec$) on each detected source, and combines the source lists from separate exposures into a single master list in order to compute mean corrected count rates.
In order to take into account the OM calibration uncertainty of the conversion factor between the count rate and the flux, we quadratically added a representative systematic error of 1.5\%\footnote{https://xmmweb.esac.esa.int/docs/documents/CAL-SRN-0378-1-1.pdf} to the statistical error of the count rate (as done in \citealt{Porquet19,Porquet24}). 

\subsection{{\sl NuSTAR} data reduction}\label{sec:nustar}

{\sl NuSTAR} \citep{Harrison13} observed ESO\,141-G55 with its two co-aligned X-ray telescopes with the corresponding focal plane modules A (FPMA) and B (FPMB) in July 2016 and October 2022 (see Table~\ref{tab:log}).
The level 1 data products were processed with the {\sl NuSTAR} Data Analysis Software (NuSTARDAS) package (version 2.1.2; February 16, 2022). Cleaned event files (level 2 data products) were produced and calibrated using standard filtering criteria with the \textsc{nupipeline} task and the calibration files available in the {\sl NuSTAR} calibration database (CALDB: 20230420).
The extraction radii for the source and background spectra were 60${\arcsec}$. The corresponding net exposure time for the observations with FPMA are reported in Table~\ref{tab:log}. The processed rmf and arf files are provided on a linear grid of 40\,eV steps.
As the full width at half maximum (FWHM)  energy resolution of {\sl NuSTAR} is 400\,eV below $\sim$50 keV and increases to 1\,keV at 86\,keV \citep{Harrison13}, we re-binned the rmf and arf files in energy and channel space by a factor of 4 to over-sample the instrumental energy resolution by at least a factor of 2.5.  
The background-corrected {\sl NuSTAR} spectra were finally binned in order to have a signal-to-noise ratio greater than four in each spectral channel. We allow for cross-calibration uncertainties between the two {\sl NuSTAR}
spectra and the {\sl XMM-Newton}-pn spectra by including in the fit a free cross-normalisation factor for the pair of {\sl NuSTAR} FPMA ($\sim$1.18) and FPMB ($\sim$1.19) spectra, with respect to the pn spectra.

\subsection{Spectral analysis method}\label{sec:method}

The {\sc xspec} (v12.12.1) software package \citep{Arnaud96} was used for spectral analysis.
We used the X-ray absorption model {\sc tbnew (version 2.3.2)} from \cite{Wilms00}, applying their interstellar medium (ISM) elemental abundances and the cross-sections from \cite{Verner96}. 
A $\chi^{2}$ minimisation is applied throughout quoting errors with 90\% confidence intervals for one interesting parameter ($\Delta\chi^{2}$=2.71). The default values of H$_{\rm 0}$=67.66\,km\,s$^{-1}$\,Mpc$^{-1}$, $\Omega_{\rm m}$=0.3111, and $\Omega_{\Lambda}=0.6889$ are assumed \citep{Planck20}.

\section{The X-ray broadband spectral analysis}\label{sec:mainX}

\begin{figure}[t!]
\begin{tabular}{cc}
\includegraphics[width=0.9\columnwidth,angle=0]{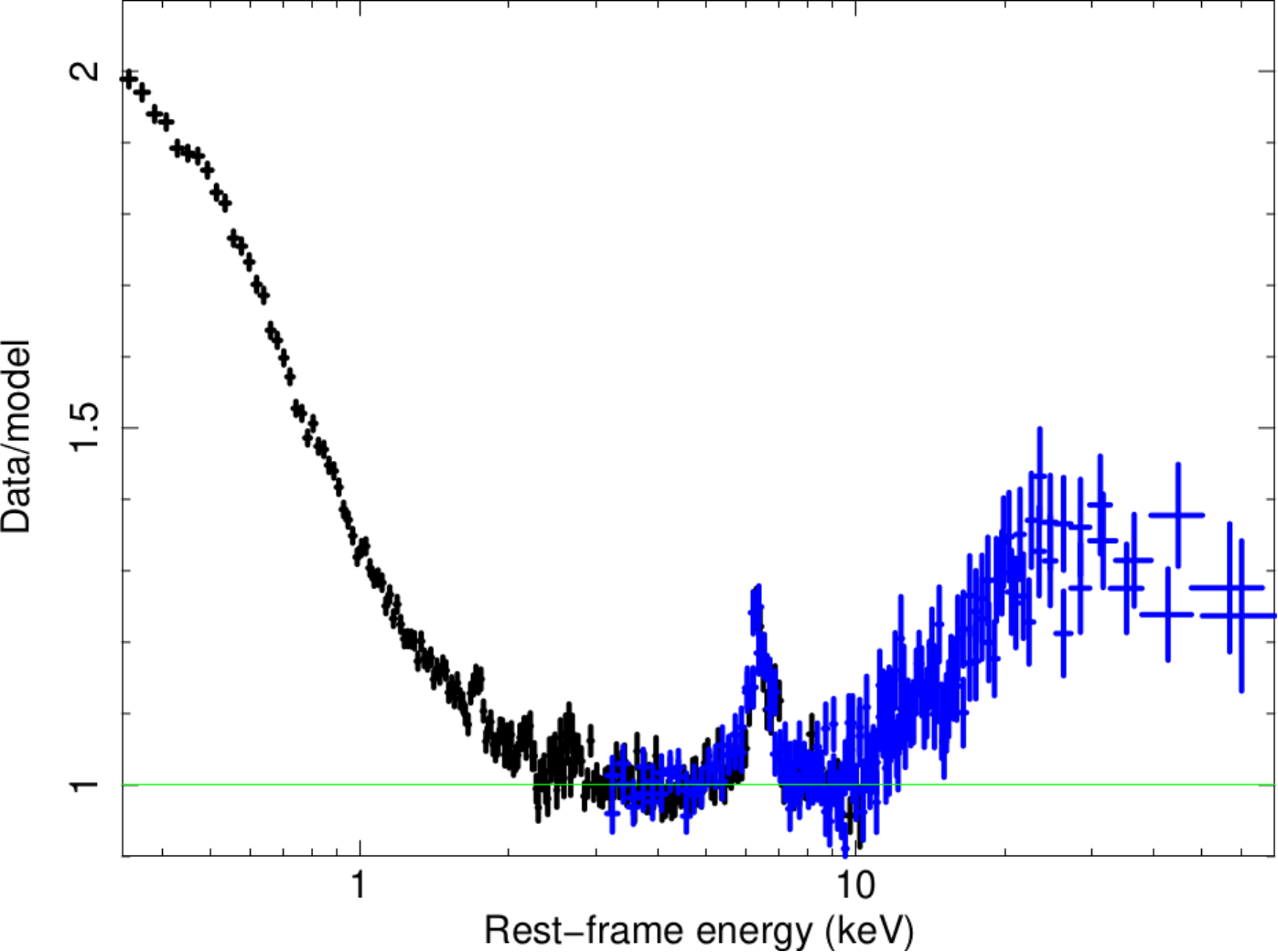}\\
\includegraphics[width=0.9\columnwidth,angle=0]{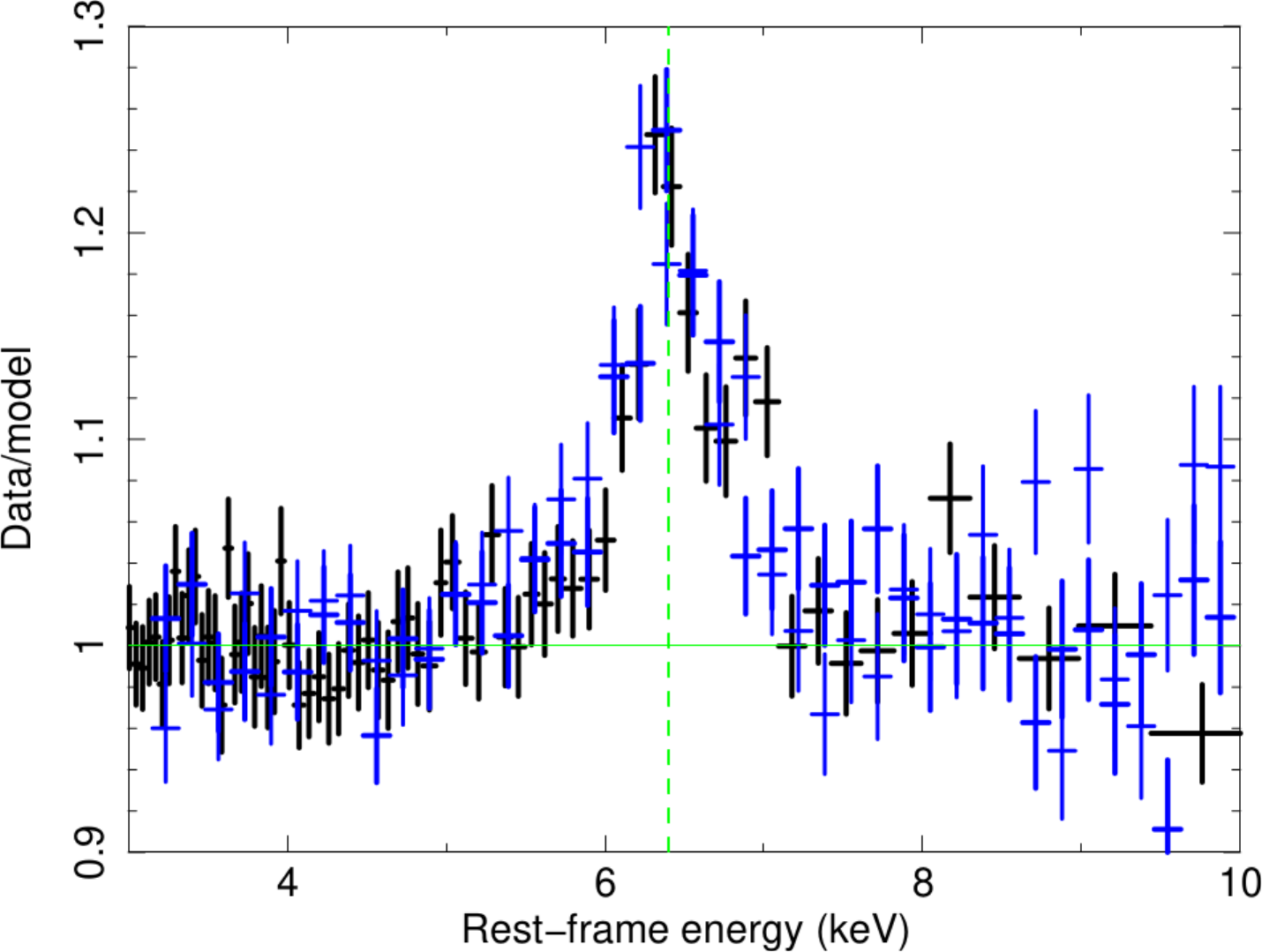}
\end{tabular}
	\caption{Data-to-model ratio of the simultaneous 2022 {\sl XMM-Newton}-pn (black) and {\sl NuSTAR} (blue) spectrum of ESO\,141-G55 fit with a Galactic (N$_{\rm H}$(Gal)=5.7$\times$10$^{20}$\,cm$^{-2}$) absorbed power-law continuum model over the 3--5\,keV energy range.
Top panel: extrapolation over the 0.3--79\,keV broadband X-ray energy range. 
Bottom panel: zoom in on the Fe\,K$\alpha$ line.
The vertical dashed green line corresponds to 6.4 keV.} 
\label{fig:3-5keV}
\end{figure}

To characterise the main X-ray components of the 2022 observation, we first fit the pn and {\sl NuSTAR} spectra between 3--5\,keV (AGN rest frame) using a power-law with Galactic absorption with $N_{\rm H}$(Gal)=5.7$\times$10$^{20}$\,cm$^{-2}$ as measured from the RGS spectral analysis (Sec.~\ref{sec:rgs}). We find a photon index of 1.93$\pm$0.03 ($\chi^{2}$/d.o.f.=119.4/119) consistent with BLS1s \citep{Porquet04a,Zhou10,Waddell20,Gliozzi20}. As illustrated in Fig.\ref{fig:3-5keV} (top panel), the fit extrapolation over the 0.3--79\,keV energy range reveal a very prominent soft X-ray excess below 2\,keV, and, in the hard X-ray range, both a significant well-resolved emission Fe\,K$\alpha$ line and a Compton hump suggesting a contribution from reflection. \\

%-----------------------------Figure --------------------------------
%-------

\subsection{Is ESO\,141-G55 still in a bare state in 2022?: the RGS data analysis}\label{sec:rgs}

\begin{figure}[t!]
\begin{tabular}{c}
\includegraphics[width=0.9\columnwidth,angle=0]{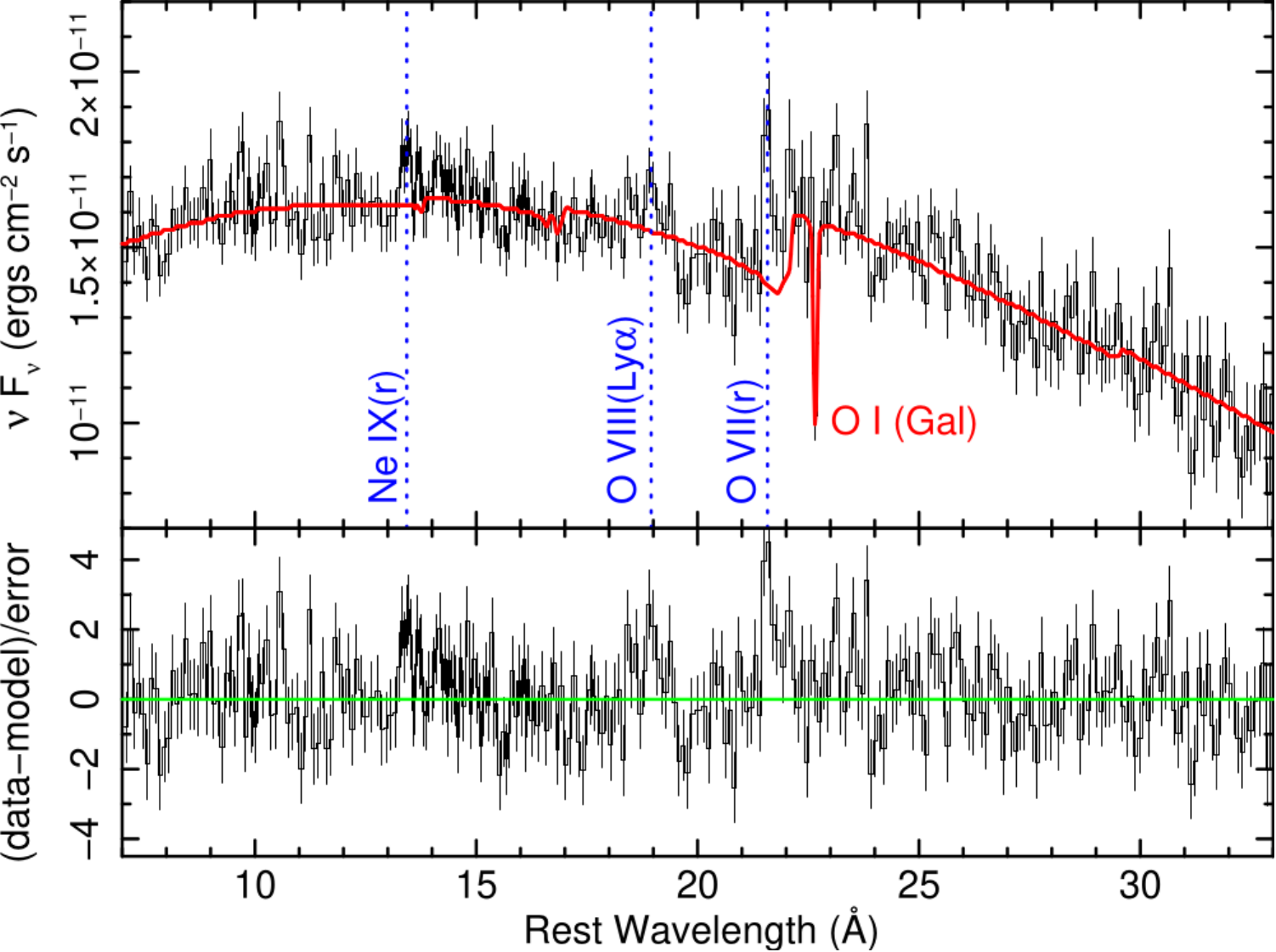}\\
\includegraphics[width=0.9\columnwidth,angle=0]{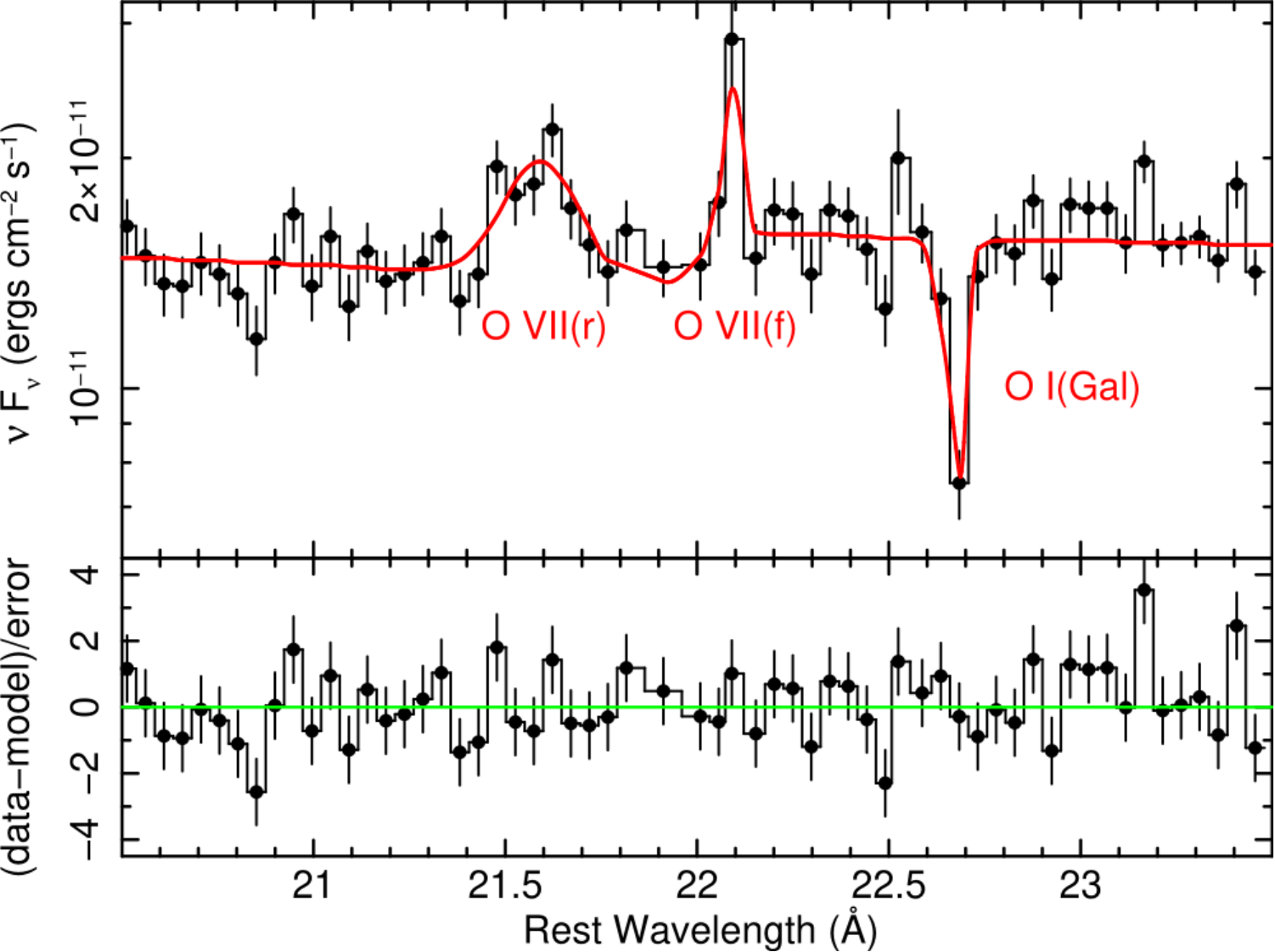}\\
\end{tabular}
	\caption{The 2022 RGS spectrum of ESO\,141-G55.
 Top panel: Fluxed spectrum over the full RGS bandpass, where the solid red line corresponds to the baseline absorbed power-law continuum, and the vertical dotted lines mark the wavelengths of the emission lines. Bottom panel: a zoom in on the \ion{O}{vii} emission band (binning of $\Delta\lambda$=0.1\AA), where the broad component at 21.6\,\AA\ corresponds to the resonance line and the narrow component at 22.1\,\AA\ corresponds to the forbidden line. The \ion{O}{i} absorption line is due to our Galaxy.}  
\label{fig:RGS}
\end{figure}

As suggested by the 2022 X-ray broadband CCD spectrum (Fig.~\ref{fig:3-5keV} top panel), ESO\,141-G55 displays an apparent smooth soft X-ray spectrum as previously reported in the 2001 and 2007 {\sl XMM-Newton} observations. Therefore, our aim is to probe any soft X-ray absorption and/or emission features at the AGN and Galactic rest frames, thanks to the high-resolution RGS spectrum.
The 2022 combined RGS spectrum was initially fitted by a simple neutral Galactic ($z$=0) absorbed power-law model. The spectrum compared to this model is shown in Fig.~\ref{fig:RGS} (top panel). The soft X-ray photon index obtained in the RGS energy band (of $\Gamma$=2.28$\pm$0.02) is much steeper compared to that obtained from the {\it XMM-Newton} and {\it NuSTAR} spectrum between 3 and 5\,keV ($\Gamma$=1.93$\pm$0.03) due to the prominent soft X-ray excess present below 2\,keV (Fig.~\ref{fig:3-5keV} top panel). The neutral Galactic absorption column was measured to be $N_{\rm H}$(Gal)=5.7$\pm$0.2$\times$10$^{20}$\,cm$^{-2}$, just slightly higher than the predicted atomic H\,\textsc{i} column of $N_{\rm H}$(Gal)=5.0$\times$10$^{20}$\,cm$^{-2}$ from 21\,cm measurements \citep{Kalberla05}, where the additional Galactic column could arise from molecular hydrogen \citep{Willingale13}. Throughout this work, N$_{\rm H}$(Gal) is then fixed to 5.7$\pm$0.2$\times$10$^{20}$\,cm$^{-2}$.\\

Although the absorbed power-law model provided an adequate description of the continuum over the RGS band, the overall fit statistic is relatively poor ($\chi^2$/d.o.f.=784.8/618). An excess of emission is present in the spectral residuals near 21.6\,\AA\ in the AGN rest frame at the $\sim$5$\sigma$ level, while two weaker excesses (at $\sim$3$\sigma$) are also present at 18.9\,$\AA$ and 13.4\,$\AA$ (Fig~\ref{fig:RGS}, upper panel). 
The rest-frame wavelengths coincide with the line emission of \ion{O}{vii} for the stronger emission at 21.6\,$\AA$, and \ion{O}{viii} and \ion{Ne}{ix} for the two weaker and shorter wavelength lines, respectively. To fit the emission, four Gaussian emission lines were added to the model, two of which are required to fit the \ion{O}{vii} triplet emission, while single Gaussian components are adequate for the likely \ion{O}{viii} and \ion{Ne}{ix} emission. The addition of the Gaussian lines significantly improved the fit statistic to $\chi^{2}$=709.6/608, while the best fit line parameters are listed in Table\,\ref{tab:RGS}. \\

\begin{table*}[t!]
  \caption{Soft X-ray emission lines in the ESO 141-G55 rest-frame observed in the 2022 RGS spectrum.}
\centering                          
\begin{tabular}{lcccccc}
\hline\hline                 
 Line ID & E\,(eV) & $\lambda$\,(\AA) & EW\,(eV) & $\sigma$\,(eV) & FWHM\,(km\,s$^{-1}$) & $\Delta\chi^2$ \\
 \hline
\ion{O}{vii} (r) & $574\pm1.1$ & $21.60\pm0.04$ & $2.4\pm0.7$ & $2.5^{+1.2}_{-1.0}$ & $3\,100^{+1500}_{-1200}$ & 36.6 \\
\ion{O}{vii} (f) & $561.5^{+2.7}_{-0.9}$ & $22.08^{+0.03}_{-0.10}$ & $0.8\pm0.5$ & $<1.0$ & $<1\,300$ & 5.7 \\
\ion{O}{viii} Ly$\alpha$ & $654.7^{+1.0}_{-0.7}$ & $18.94^{+0.02}_{-0.03}$ & $0.8\pm0.3$& $<2.5$ & $<2\,700$ & 14.0 \\
\ion{Ne}{ix} (r) & $923.0^{+4.0}_{-3.2}$ & $13.43\pm0.05$ & $2.2\pm0.8$ & $5.3^{+3.5}_{-1.7}$ & $4\,000^{+2700}_{-1300}$ & 18.9\\
\hline
\hline    
\hline                  
\end{tabular}
\label{tab:RGS}
\end{table*}

%-----------------------------Figure --------------------------------
%-------

Figure~\ref{fig:RGS} (bottom panel) shows a zoom in around the \ion{O}{vii} emission band, illustrating the two possible components of emission. The strongest feature statistically speaking ($\Delta$$\chi^2$=36.6 for three additional parameters) occurs at 21.60$\pm$0.04\,\AA\  (574.1$\pm$1.1\,eV), corresponding to the expected laboratory frame wavelength of the \ion{O}{vii} resonance line \citep[e.g.,][]{Porquet01}. This line is broadened, with a fitted width of $\sigma$=2.5$^{+1.2}_{-1.0}$\,eV, corresponding to a FWHM velocity width of 3100$^{+1500}_{-1200}$\,km\,s$^{-1}$, which is consistent with the H$\alpha$, H$\beta$ and Ly$\alpha$ line widths of 3010\,km\,s$^{-1}$, 3730\,km\,s$^{-1}$ \citep{Stirpe90}, and 3500\,km\,s$^{-1}$ \citep{Turler98}, respectively, emitted by the broad line region (BLR), but slightly smaller than the \ion{C}{iv} line width (5250\,km\,s$^{-1}$;\citealt{Turler98}). 

The second \ion{O}{vii} emission component occurs at 22.1$\pm$0.1\,\AA\ (561.5\,eV) and appears to be narrow, arising from just a single resolution bin in the RGS spectrum (see Fig.~\ref{fig:RGS}, bottom panel). It is statistically weaker (with $\Delta\chi^2$=5.7 for two additional parameters). If confirmed, this is likely to be associated with the emission of the forbidden line at an expected wavelength of 22.1\,$\AA$. As the line is not resolved, only an upper limit can be placed on its width, of $\sigma$$<$1\,eV. Moreover, the lack of intercombination line would point out to an emitting region with a density lower than about 10$^{9}$\,cm$^{-3}$, which corresponds to the critical density value for \ion{O}{vii} \citep{Porquet00, Porquet10}. Therefore, this narrow \ion{O}{vii} emission component could be associated with a farther out region such as the narrow line region (NLR).
Finally, we note that the Ne\,\textsc{ix} emission is broadened (see Table~\ref{tab:RGS}), with a velocity width consistent with the \ion{O}{vii} line (although less well determined) and its centroid wavelength is also consistent with the resonance line emission. In contrast, the H-like \ion{O}{viii} line appears unresolved. No other emission lines are detected in the RGS spectrum. \\

The RGS spectrum does not show strong absorption lines against the continuum, apart from the expected O\,\textsc{i} neutral absorption line associated with our Galaxy (e.g., Fig.~\ref{fig:RGS} bottom panel). Nonetheless, we attempted to place limits on the column and ionisation of any warm absorbing gas associated with ESO\,141-G55 or its host galaxy. To this end, an \textsc{xstar} \citep{Kallman01} multiplicative photoionised absorption table was added to the model, adopting a host galaxy redshift of z=0.0371 and covering a range of ionisation from $\log\xi$=-1.5--3.5 and for a turbulence velocity of 200\,km\,s$^{-1}$. The absorption column decreased to the lowest allowed column calculated in the \textsc{xstar} grid (of $N_{\rm H}$=8$\times10^{19}$\,cm$^{-2}$) and the highest possible ionisation (log\,$\xi$=3.5). Such an absorber does not produce any measurable absorption lines in the RGS spectrum or improve the fit statistic, given the very low opacity of the absorber. Formally, an upper limit to the column of $N_{\rm H}$$<$4$\times$10$^{20}$\,cm$^{-2}$ was obtained and a corresponding lower limit to the ionisation parameter of $\log$\,$\xi$$>$2.9 was also obtained. Thus, there is no significant warm-absorbing gas in our line of sight towards this AGN, consistent with its classification as a bare Seyfert\,1 galaxy.

\subsection{Characterisation of the Fe\,K$\alpha$ emission complex}\label{sec:FeK}

\begin{figure}[t!]
\begin{tabular}{c}
\includegraphics[width=0.9\columnwidth,angle=0]{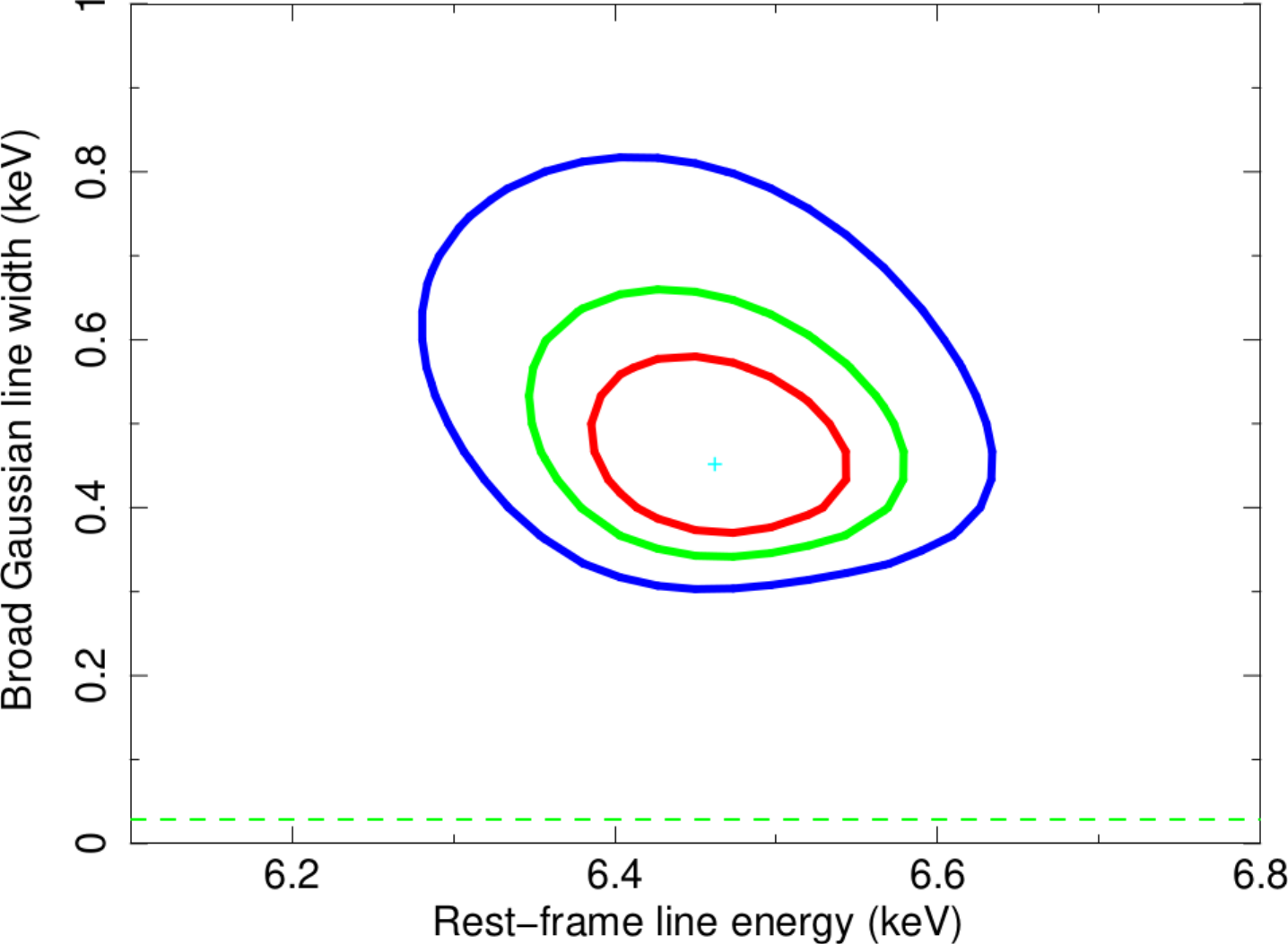}\\
\end{tabular}
	\caption{Contour plot of the broad Gaussian line width (keV) versus its rest-frame energy (keV). The red, green and blues curves show the confidence levels at 68\% ($\Delta\chi^2$=2.3), 90\% ($\Delta\chi^2$=4.61) and 99\% ($\Delta\chi^2$=9.21), respectively. The horizontal dashed green line corresponds to a FWHM width of 
 $\sim$3000\,km\,s$^{-1}$ (BLR).}
\label{fig:FeKcontour}
\end{figure}

To determine the origin of the Fe\,K$\alpha$ complex (Fig.~\ref{fig:3-5keV} bottom panel), we fit the 2022 pn spectrum between 3 and 10\,keV with a Galactic absorbed power-law adding two Gaussian lines: one broad component with its energy and width allowed to vary as the main profile is well resolved, and one narrow unresolved core component (E=6.40\,keV,  $\sigma$=0\,eV) to account for possible neutral reflection from the outer BLR and/or molecular torus as suggested by the Fe\,K peak line profile.
The narrow core component is found to have a low equivalent width (EW) of 30$\pm$10\,eV, while the broad component (at 6.46$\pm$0.09\,keV) has a much larger EW of 140$\pm$30\,eV ($\chi^{2}$/d.o.f.=382.9/343). The broad line width is $\sigma$=450$^{+140}_{-90}$\,eV  corresponding to a full width at half maximum (FWHM) of 50000$^{+15000}_{-10000}$\,km\,s$^{-1}$. This line component is much broader than than expected from the BLR ($\sim$3000\,km\,s$^{-1}$, Sec.~\ref{sec:rgs}) pointing out that it originates from the accretion disc (Fig.~\ref{fig:FeKcontour}). Therefore, we replace the broad Gaussian line with a relativistic line profile using {\sc relline} \citep{Dauser10}. The black hole spin was fixed to zero since otherwise it was not constrained by the model. We fixed the disc emissivity index value ($q$; with emissivity $\propto R^{-q}$) to the canonical value of three, and allowed the inner radius of relativistic reflection to vary, that is to say, $R_{\rm in}$ is no longer set by default to the inner stable circular orbit (ISCO). The line EWs are 180$\pm$30\,eV and 41$\pm$9\,eV for the relativistic line and the narrow line, respectively ($\chi^{2}$/d.o.f=374.9/343). We measure a disc inclination of $\theta$=43$^{+5}_{-2}$\,degrees  consistent with the value inferred from accretion disc fits to the UV continuum and the H$\beta$ emission line for a non-spinning black hole \citep{Rokaki99}.
The inner radius value inferred from the relativistic reflection component, $R_{\rm in}$=11$^{+9}_{+3}$\,$R_{\rm g}$, was found to be insensitive to the black hole spin value, which indicates that relativistic reflection occurs in the inner part of the accretion disc but not down to ISCO.

%-----------------------------Figure --------------------------------
%-------

\subsection{Search for a UFO signature and comparison with the 2007 {\sl XMM-Newton} observations}\label{sec:UFO}

As previously mentioned, a significant possible wind signature for ESO\,141-G55 was reported on October 30 2007 (2007\#4). We then attempt to place limits on the presence of any UFO signature located in the Fe\,K$\alpha$ energy band during the 2022 {\sl XMM-Newton} observation. 
The 2022 pn spectrum was fitted with a Galactic absorbed power-law by adding an {\sl xstar} absorption table with a turbulence velocity of 1000\,km\,s$^{-1}$. The outflow velocity ($v_{\rm out}$) was allowed to vary between 0 and 0.3\,$c$ and the ionisation parameter (log\,$\xi$) was varied over the range from 3 to 5, where these values are typical of other UFOs from photoionisation modelling \citep[e.g.][]{Tombesi10,Gofford15}. The addition of a fast absorber did not improve the fit statistics and only upper-limits were found. For example, an upper limit to the column density of $N_{\rm H}$$<$7$\times$10$^{21}$\,cm$^{-2}$ was found for an ionisation of $\log$\,$\xi$=3.5, with somewhat higher column densities allowed for a higher ionisation parameter, e.g. for $\log$\,$\xi$=4, then $N_{\rm H}$$<$2$\times$10$^{22}$\,cm$^{-2}$. Therefore, there is no evidence in this 2022 very high signal-to-noise ratio X-ray spectrum of ESO\,141-G55 of the presence of UFOs.\\

\begin{figure}[t!]
\begin{tabular}{c}
\includegraphics[width=0.9\columnwidth,angle=0]{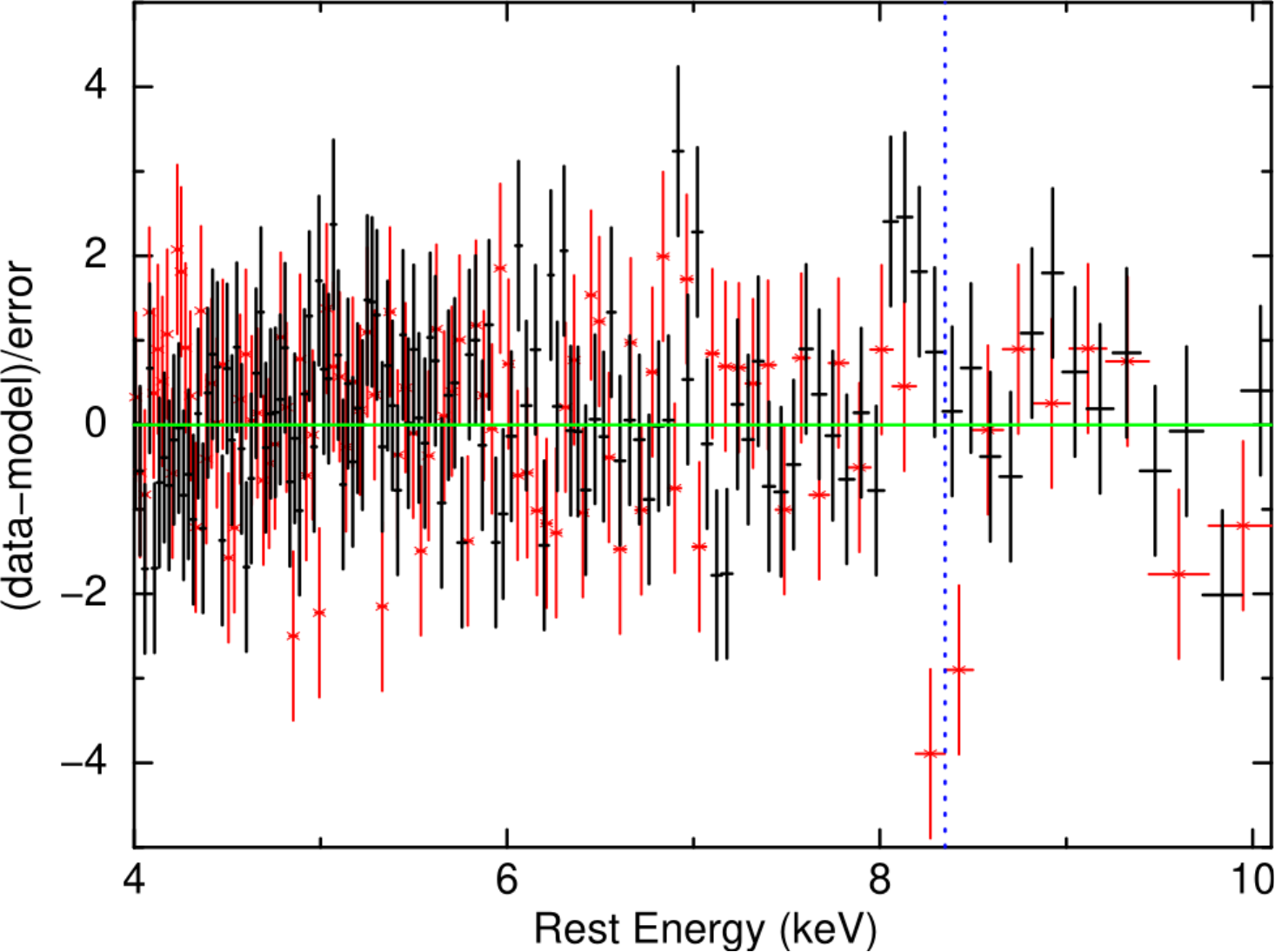} \\
\end{tabular}
\caption{Fit of the 2022 (black) and 2007\#4 (red) 3-10\,keV {\sl XMM-Newton}-pn observations with a Galactic power-law plus two  Gaussian lines: one broad and one narrow (see text). The 2007\#4 spectrum displays an absorption feature (dotted line) at about 4$\sigma$ below the continuum, while there is nothing apparent at a similar energy in the 2022 observation. }
\label{fig:UFO}
\end{figure}

To provide a direct comparison between the present 2022 and the 2007\#4 observations, we reprocessed the 2007\#4 pn spectra using the same calibration as used for the present 2022 observation, but using an annular extraction region to minimise the pile-up effect (see Sect.~\ref{sec:xmm} for more details). The 2022 and 2007\#4 pn observations were fitted over the 3--10 keV band with a Galactic absorbed power-law adding two Gaussian (broad and narrow) line components, with the same parameters as for the 2022 spectrum, to fit the Fe K$\alpha$ emission. Figure~\ref{fig:UFO} displays the residuals of both spectra, where the position of the absorption line is marked by the dotted line. This confirms that, as previously found by \cite{DeMarco13}, the 2007 absorption feature stands out at about 4\,$\sigma$ below the continuum, while there is nothing apparent at a similar energy in the 2022 observation. When fitting the absorption feature with a narrow Gaussian ($\sigma$=10\,eV) of negative normalisation,  an equivalent width of EW=77$\pm$28\,eV is found in the 2007\#4 spectrum and a very small upper limit of $<$6\,eV in the 2022 data assuming that it occurs at the same rest frame energy of 8.34$\pm$0.05\,keV. The improvement in fit statistic for the 2007\#4 feature is $\Delta$$\chi^2$=23.4. For the combined pn spectrum of the other three shorter 2007 observations, performed between October 9 and 12, 2007 (Table~\ref{tab:log}), an upper limit of the EW of $<$48 eV is measured for the absorption feature assuming the same centroid energy as above. This is just barely inconsistent with the detection in the 2007\#4 pn spectrum at the 90\% confidence level.
The absorption feature present in the 2007\#4 spectrum was then modelled with the {\sc xstar} absorber applied previously to the 2022 spectrum. The best fit of the UFO properties are  N$_{\rm H}$=3.8$^{+3.5}_{-1.9}$$\times$10$^{23}$\,cm$^{-2}$, log\,$\xi$=4.5$^{+0.5}_{-0.2}$ and $v_{\rm out}$=-0.176$\pm$0.006\,c.
In the case of ESO\,141-G55, the absorption arises purely from $\ion{Fe}{xxvi}$. Compared to the same velocity and ionisation as above, the upper limit to the column density for the 2022 observation is $N_{\rm H}$$<$1.7$\times$10$^{22}$\,cm$^{-2}$, then about an order of magnitude below what was found in the 2007\#4 observation. 

\subsection{The very hot corona and relativistic reflection properties at two epochs (2022 and 2016)}\label{sec:above3keV}

%---------- TABLE ---------
\begin{table}[t!]
  \caption{Simultaneous 3--79\,keV fits, using the {\sc reflkerrd} relativistic reflection model, of the October 2022 {\sl XMM-Newton+NuSTAR} and July 2016 {\sl NuSTAR} observations of ESO\,141-G55 (see Sect.~\ref{sec:above3keV} for details).
}
\centering                          
\begin{tabular}{@{}l l l }
\hline\hline                 
 Parameters               &   \multicolumn{1}{c}{2022}   &   \multicolumn{1}{c}{2016} \\
                          &   \multicolumn{1}{c}{\sl NuSTAR} &   \multicolumn{1}{c}{\sl NuSTAR}\\
                          &   \multicolumn{1}{c}{+{\sl XMM}} & \\
 
 \hline
\hline   
$\theta$ (deg) & \multicolumn{2}{c}{44$^{+4}_{-2}$} \\
$kT_{\rm hot}$ (keV)                 &  136$^{+8}_{-18}$               & 124$^{+36}_{-19}$          \\
$\tau_{\rm hot}$                     & 0.31$^{+0.05}_{-0.03}$           & 0.41$^{+0.23}_{-0.09}$               \\
$q$           & \multicolumn{2}{c}{2.0$\pm$0.3}                  \\
log\,$\xi$ (erg\,cm\,s$^{-1}$)       &   \multicolumn{2}{c}{1.7$^{+0.1}_{-0.2}$}     \\
$\cal{R}$                            & 0.76$^{+0.09}_{-0.10}$  &  0.52$^{+0.11}_{-0.05}$        \\
$A_{\rm Fe}$ & \multicolumn{2}{c}{1.7$^{+0.1}_{-0.2}$} \\
$norm_{\rm reflkerrd}$ ($\times$10$^{-3}$) &  9.2$^{+0.1}_{-0.3}$     & 9.1$^{+0.5}_{-1.0}$     \\
$\chi^{2}$/d.o.f.  ($\chi^2_{\rm red}$) & \multicolumn{2}{c}{1256.1/1194 (1.05)}  \\
\hline
$F^{\rm unabs}_{\rm 3-79\,keV}$$^{(a)}$ (erg\,cm$^{-2}$\,s$^{-1}$) &  8.0$\times$10$^{-11}$ & 7.2$\times$10$^{-11}$ \\
$L^{\rm unabs}_{\rm 3-79\,keV}$$^{(a)}$ (erg\,s$^{-1}$) & 2.7$\times$10$^{44}$ & 2.5$\times$10$^{44}$ \\
\hline    
\hline                  
\end{tabular}
\label{tab:reflkerrDabove3keV}
\flushleft
\small{{\bf Notes}. $^{(a)}$ Source flux and luminosity corrected from Galactic absorption.}
\end{table}

\begin{figure}[t!]
\begin{tabular}{c}
 \includegraphics[width=0.9\columnwidth,angle=0]{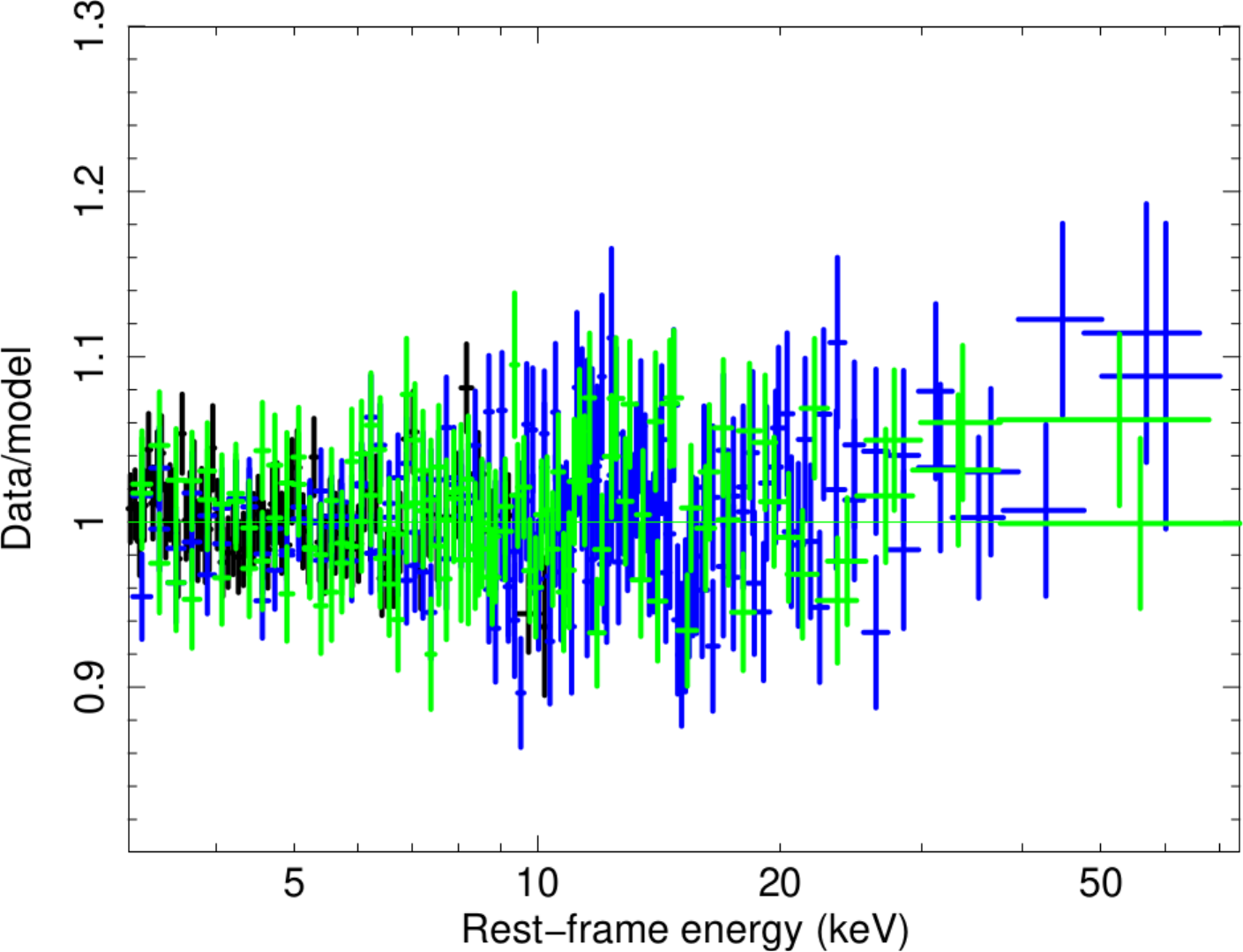} \\
\end{tabular}
\caption{Data-to-model ratio of the fit above 3 keV of the 2022 simultaneous XMM-Newton (black) and NuSTAR (blue) spectrum and of the 2016 NuSTAR (green) spectrum using the {\sc reflkerrd} relativistic reflection model with a primary Comptonisation continuum shape.}
\label{fig:reflkerrDabove3keV}
\end{figure}
%-----------------

To determine the hot corona and relativistic reflection properties of ESO\,141-G55, we focus on its hard X-ray shape using the simultaneous 2022 {\sl XMM-Newton} and {\sl NuSTAR} data above 3\,keV to prevent the fit from being driven by the soft X-ray emission.
 We also include in this study the previous {\sl NuSTAR} observation of ESO\,141-G55 performed in 2016 \citep[Table~\ref{tab:log}; ][]{Ezhikode20,Panagiotou20,Akylas21,Kang22} to check for any variation in the temperature of the hot corona between the two epochs.\\

A simple power-law ($\Gamma$) continuum with an exponential cutoff at high energy ($E_{\rm cut}$) and a broad Gaussian line cannot reproduce the Compton hump shape due to relativistic reflection on the accretion disc ($\Delta$$\chi^2$/d.o.f.=1724.0/1197).
Therefore, we consider the physical model {\sc reflkerrd}\footnote{The usage notes as well as the full description of the models and their associated parameters are reported at \href{https://users.camk.edu.pl/mitsza/reflkerr/reflkerr.pdf}{https://users.camk.edu.pl/mitsza/reflkerr/reflkerr.pdf}} \citep[version 2019;][]{Niedzwiecki19} which includes both the hot corona and relativistic reflection components. The reflection fraction $\cal{R}$ is defined as the amount of reflection $\Omega$/(2*$\pi$). We use a primary Comptonisation continuum shape, which is more physical and has a sharper high-energy roll-over compared to an exponential cutoff power-law. Furthermore, such a model has the advantage of directly measuring the hot corona temperature ($kT_{\rm hot}$). We note that previously only the cutoff energy was measured using the 2016 {\sl NuSTAR} observation \citep{Ezhikode20,Panagiotou20,Akylas21,Kang22}. The hard X-ray shape of the reflected component of {\sc reflkerrd} is calculated using {\sc ireflect} (the xspec implementation of the exact model for Compton reflection of \citealt{Magdziarz95} in the hard X-ray range) convolved with {\sc compps}. The {\sc compps} model \citep{Poutanen96} appears to be a better description of thermal Comptonisation than {\sc nthcomp} \citep{Zdziarski96,Zycki99} when compared to Monte Carlo simulations \citep{Zdziarski20}. In appendix~\ref{app:relxill}, we report for comparison the fit results for the {\sc relxillcp} models for which the hard Comptonisation (hot corona) X-ray shape is instead calculated using the {\sc nthcomp} model.\\

We assume a slab geometry for the hot corona with a single power-law disc emissivity index ($q$; with emissivity $\propto R^{-q}$) allowed to vary. Indeed, a slab-like geometry seems to be favoured by X-ray spectro-polarimetry observations of AGN by the {\sl Imaging X-ray Polarimetry Explorer} \citep[IXPE; e.g.,][]{Gianolli23}. However assuming a lamppost geometry gives similar results for the present study (Appendix~\ref{app:relxill}). Here we allowed the disc emissivity index to vary and set the radius of the inner disc to the ISCO, since otherwise it is not constrained due to the degeneracy between both parameters. This emissivity index value will then be compared below to that required when fitting the whole 0.3–79 keV energy range. Since the spin value is not constrained, it was fixed at zero. We checked that assuming other spin values did not impact our results. The temperature of the thermal seed photons ($kT_{\rm bb}$) Comptonised by the hot corona is an explicit physical parameter of this model. We fixed it to 10\,eV corresponding to the expected maximum temperature of the accretion disc around a black hole mass of $\sim$1.3$\times$10$^{8}$\,M$_{\odot}$ accreting at a $\sim$10\% Eddington rate. To take into account the unresolved (weak) core of the Fe\,K$\alpha$ complex, a narrow Gaussian line was added at 6.4\,keV as done in Sect.~\ref{sec:FeK}. We note that similar parameter values for the {\sc reflkerrd} model are inferred if a molecular torus model, such as {\sc borus12} \citep{Balokovic18,Balokovic19}, is used instead of a narrow Gaussian line. Indeed, the narrow core of the Fe\,K$\alpha$ component is weak for ESO\,141-G55 meaning that any associated reflection contribution to the hard X-ray spectrum is negligible.
We tied between both epochs the inclination angle and the iron abundance of the disc, which are not supposed to vary on year timescales. Moreover, the emissivity indices between both epochs are tied since they are found to be compatible.
A good fit is found (Fig.~\ref{fig:reflkerrDabove3keV}) with the best-fit parameters reported in Table~\ref{tab:reflkerrDabove3keV}.
Although the main value of the reflection fraction is higher in 2022, the increase is only at a 1.6$\sigma$ confidence level when taking into account the error bars. 
The hot corona has consistent temperatures at both epochs ($kT_{\rm hot}$$\sim$120--140\,keV) within their error bars and is found to be optically thin ($\tau_{\rm hot}$$\sim$0.3--0.4). \\

\begin{figure}[t!]
\begin{tabular}{c}
\includegraphics[width=0.9\columnwidth,angle=0]{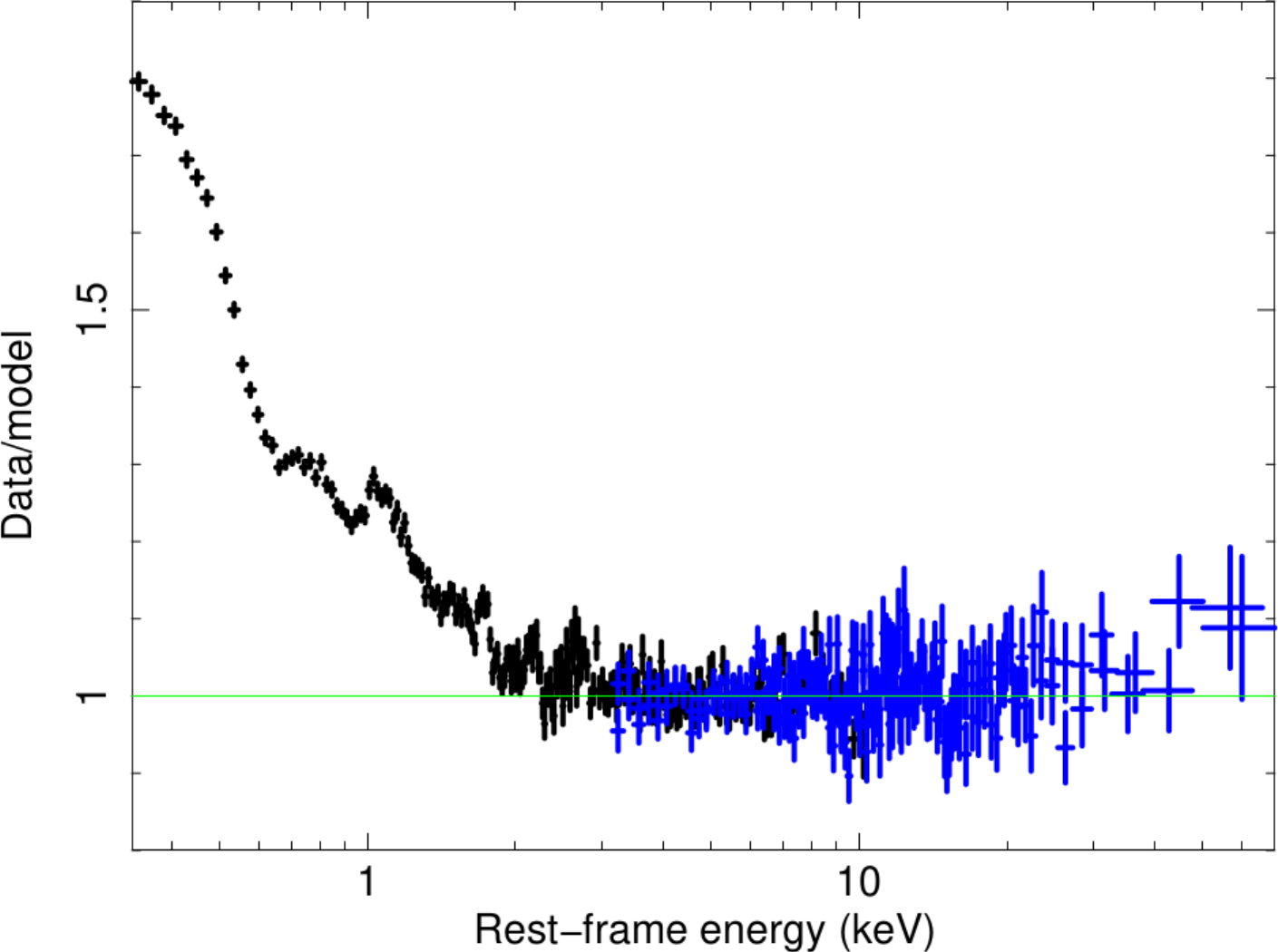} \\
\includegraphics[width=0.9\columnwidth,angle=0]{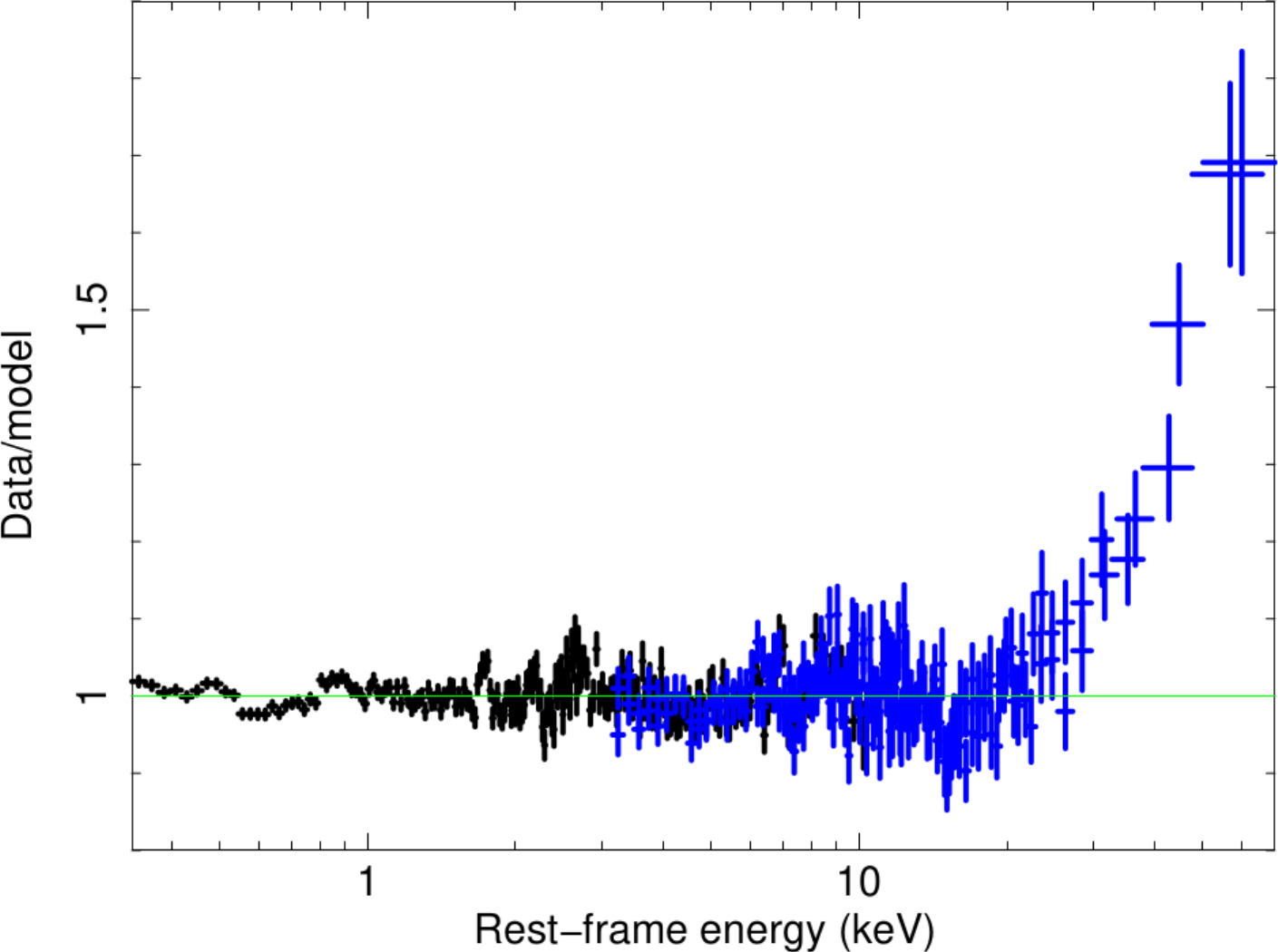}\\
\end{tabular}
\caption{Fits of the 2022 simultaneous {\sl XMM-Newton}-pn (black) and {\sl NuSTAR} (blue) spectrum of ESO\,141-G55 with the {\sc reflkerrd} model. Top panel: fit above 3\,keV (Fig.~\ref{fig:reflkerrDabove3keV}, Table~\ref{tab:reflkerrDabove3keV}), which has been extrapolated down to 0.3\,keV.  Bottom panel: fit over the entire 0.3--79\,keV energy range, here allowing for a broken power-law emissivity, and a spin value and a disc density free to vary.
  }
\label{fig:reflkerrD}
\end{figure}

The extrapolation of the 2022 fit down to 0.3\,keV shows that the soft X-ray excess is not accounted for with relativistic reflection alone, leaving a huge positive residual below $\sim$2\,keV (Fig.~\ref{fig:reflkerrD} top panel, $\chi^2$/d.o.f=96229/924). We notice, that due to the lack of simultaneous XMM-Newton data to the 2016 {\sl NuSTAR} data we cannot perform such an extrapolation.
When this model is applied to the whole 0.3--79\,keV X-ray range of the 2022 spectrum, here allowing for a broken power-law emissivity\footnote{In the case of a broken power-law emissivity, q$_{1}$ is the emissivity index at $r$$<$$R_{\rm br}$ and q$_{2}$ is the emissivity index at $r$$>$$R_{\rm br}$. $R_{\rm br}$, the breaking radius, is expressed in gravitational radii.}, for a spin value and a disc density free to vary, though the $\chi^2$ value drastically decreases ($\chi^2$/d.o.f=1436.4/920), it fails to both reproduce the soft and hard X-ray shapes (Fig.~\ref{fig:reflkerrD} lower panel). Indeed, the broadband fit is driven by the soft X-ray data for which a very steep disc emissivity, q$\sim$7.8 below about 8\,$R_{\rm g}$ and a much higher reflection fraction $\cal{R}$$\sim$4.3, would be required by relativistic reflection alone to reproduce it, contrary to the hard X-ray excess requiring q$\sim$2.0 and $\cal{R}$$\sim$0.8.
We note that similar results are found for the {\sc relxillcp} model \citep{Dauser13}, as reported in the Appendix~\ref{app:relxill}.

%####################################################

\section{Disc-corona properties: SED (UV to hard X-rays) spectral analysis using the {\sc relagn} model}\label{sec:SED2022}

As shown in the previous section, the relativistic reflection is unable to reproduce both the soft X-ray excess and the hard X-ray shape. Therefore, we explore the hybrid scenario where Comptonisation from both a warm and hot corona and relativistic reflection are present, as found for Ark\,120 \citep{Matt14,Porquet18,Porquet19}, Fairall\,9 \citep{Lohfink16}, TON\,S180 \citep{Matzeu20}, and Mrk\,110 \citep{Porquet21,Porquet24}. \\ 

\begin{figure}[t!]
\includegraphics[width=0.9\columnwidth,angle=0]{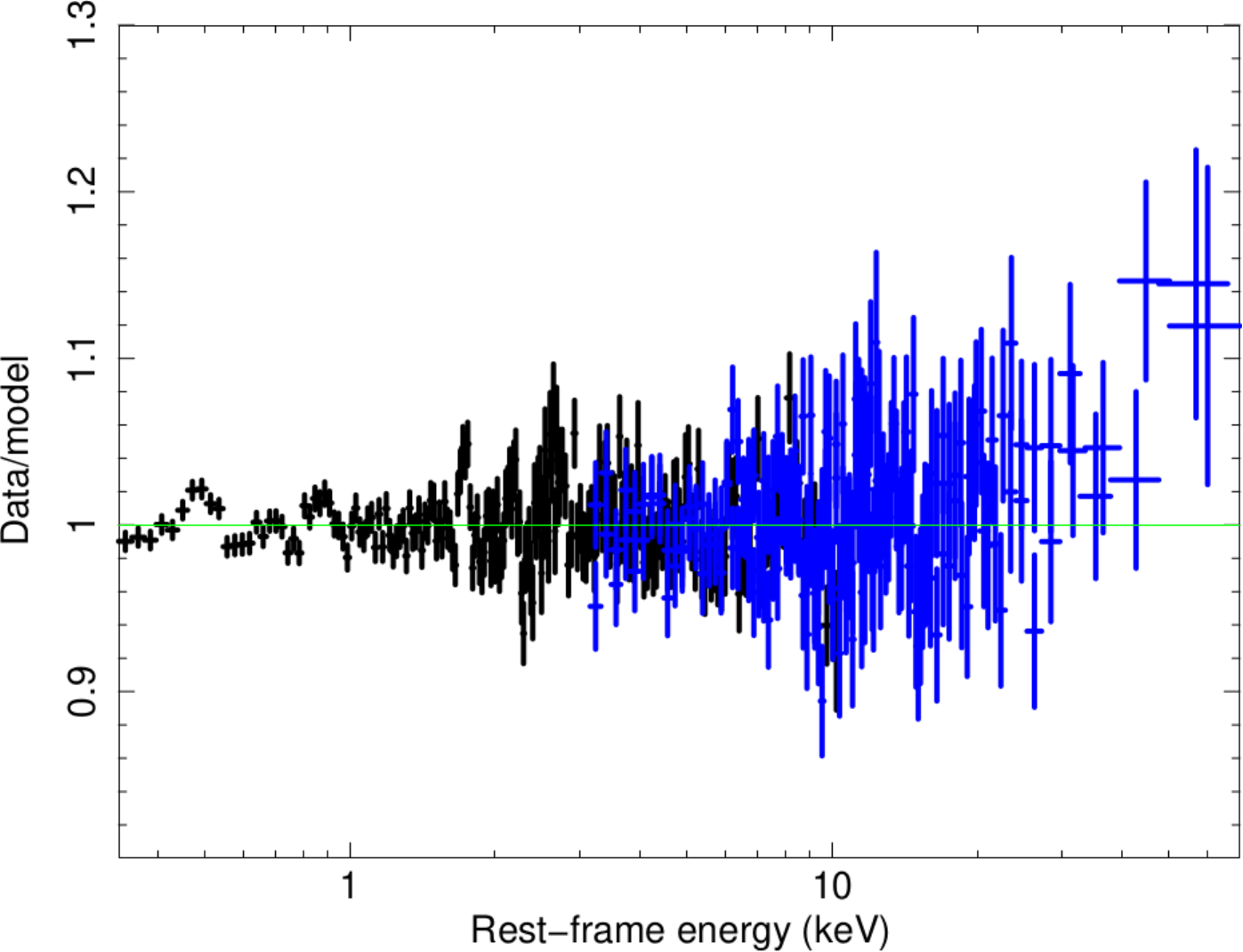} \\
\caption{Fit of the 2022 0.3-79\,keV simultaneous {\sl XMM-Newton}-pn (black) and {\sl NuSTAR} (blue) spectrum of ESO\,141-G55 using a simplistic hybrid model combining the {\sc comptt} (warm corona) and {\sc reflkerrd} (hot corona and relativistic reflection) models (see Sect.\ref{sec:SED2022} for details).}
\label{fig:comptt}
\end{figure}

As a very first step, we test whether the soft X-ray excess can originate from a warm corona using a simple modelling with the {\sc comptt} model \citep{Titarchuk94} assuming a slab geometry. We still use the {\sc reflkerrd} model to account for the emission from both the hot corona and the relativistic reflection (Sect.~\ref{sec:above3keV}). For the relativistic reflection component, we assumed a canonical value of three for the emissivity index, but we allowed the inner radius (expressed in R$_{\rm g}$) to be free to vary. As the spin value was not constrained by this modelling, it was set to zero, without a noticeable impact on the results.  
As illustrated in Fig.~\ref{fig:comptt}, this very simple modelling can broadly fit the X-ray spectrum, indicating that an optically thick warm corona (with $kT_{\rm warm}$$\sim$0.3\,keV and $\tau_{\rm warm}$$\sim$12) could indeed explain the prominent soft X-ray excess in ESO\,141-G55.  The best fit parameters are reported in Table~\ref{tab:comptt}. The inner radius of the relativistic reflection of about 20\,$R_{\rm g}$, consistent within the error bars with the value found from the Fe\,K$\alpha$ line study (Sect.~\ref{sec:FeK}), confirms that it occurs in the inner part of the disc-corona system but not down to ISCO. \\

%---------- TABLE ---------
\begin{table}[t!]
  \caption{0.3--79\,keV fit of the 2022 simultaneous {\sl XMM-Newton/NuSTAR} observation of ESO\,141-G55, using {\sc comptt} (warm corona) and {\sc reflkerrd} (hot corona and relativistic reflection) models (see Sect.~\ref{sec:SED2022} for details).}
\centering                          
\begin{tabular}{@{}l l  }
\hline\hline                 
 Parameters               &   \\
 
 \hline
\hline   
 &   {\sc comptt}  \\
\hline
$kT_{\rm warm}$ (keV)                 &  0.33$^{+0.03}_{-0.01}$           \\
$\tau_{\rm warm}$                     & 11.5$\pm$0.1           \\
$norm_{\rm comptt}$  &    5.9$^{+1.1}_{-0.1}$    \\
\hline
 & {\sc reflkerrd} \\
\hline
$kT_{\rm hot}$ (keV)                 &  134$^{+4}_{-2}$               \\
$\tau_{\rm hot}$                     & 0.31$^{+0.03}_{-0.01}$          \\
$\theta$ (deg) & 43$^{+2}_{-1}$ \\
$a$    & 0 (f) \\
$q$           &  3.0 (f)                  \\
$R_{\rm in}$ (R$_{\rm g}$) &   21$^{+6}_{-5}$     \\
log\,$\xi$ (erg\,cm\,s$^{-1}$)       &   0.3$\pm$0.1     \\
$\cal{R}$                            & 0.79$^{+0.02}_{-0.03}$  \\
$A_{\rm Fe}$ & 1.0$^{+0.2}_{-0.1}$ \\
$norm_{\rm reflkerrd}$ ($\times$10$^{-3}$) & 9.4$\pm$0.1         \\
\hline
$\chi^{2}$/d.o.f.  & 1143.6/921  \\
\hline
$F^{\rm unabs}_{\rm 0.3-79\,keV}$ (erg\,cm$^{-2}$\,s$^{-1}$) &  1.4$\times$10$^{-10}$ \\
$L^{\rm unabs}_{\rm 0.3-79\,keV}$ (erg\,s$^{-1}$) & 4.7$\times$10$^{44}$ \\
\hline    
\hline                  
\end{tabular}
\label{tab:comptt}
\flushleft
\small{{\bf Notes}. $^{(a)}$ Source flux and luminosity corrected from Galactic absorption.}
\end{table}

We now aim to determine with a more physical approach the properties of the disc-corona system of ESO\,141-G55 using its spectral energy distribution (SED) from UV to hard X-rays. 
For this, we use the new {\sc relagn} model, which is based on the {\sc agnsed} code of \cite{Kubota18}, but which incorporates general relativistic ray-tracing \citep{Hagen23b}. The model consists of an inner hot corona ({\bf $R_{\rm ISCO}$}$\leq$$R$$\leq$\,$R_{\rm hot}$), a warm Comptonised disc ($R_{\rm hot}$\,$\leq$$R$$\leq$\,$R_{\rm warm}$) and an outer standard disc ($R_{\rm warm}$$\leq$$R$$\leq${\bf $R_{\rm out}$}). An illustration of this disc-corona geometry is displayed in Fig.~2 from \cite{Kubota18}. 
The model parameters of {\sc relagn} are identical to those of {\sc agnsed}, except an additional parameter that allows a colour temperature correction to the standard outer disc ($f_{\rm col}$). In {\sc agnsed}, the colour temperature correction is hardwired at unity.  A detailed description of the {\sc relagn} model is provided in \cite{Hagen23b}.
Since as shown in the previous section a relativistic reflection is definitively present - but not included in the {\sc relagn} model - we also added the {\sc reflkerrd} model. The emissivity indices are both fixed to the canonical value  of three. The inner radius of the relativistic reflection component is set to the $R_{\rm hot}$ radius since the disc is truncated below this value. We also included the contribution of the weak narrow core of the Fe\,K$\alpha$ line using a narrow Gaussian line as done in Sect.~\ref{sec:FeK}. Our baseline model is: {\sc tbnew$\times$redden(relagn + reflkerrd + zgaussian)}.\\

 \begin{table}[t!]
\caption{Best-fit result of the simultaneous 2022 SED (UV to hard X-rays) of ESO\,141-G55
 with the {\sc relagn+reflkerrd} baseline model (see Sect.~\ref{sec:SED2022} for details).}
 %See text for detailed explanations.}
\centering
\begin{tabular}{@{}l c }
\hline\hline
%parameter & \multicolumn{1}{c}{2022}  \\
%\hline
$a$ & $\leq$0.2 \\
$\theta$ (degrees) & 43$^{+2}_{-3}$\\
log\,$\dot{m}$ & $-$0.93$\pm$0.01 \\
$kT_{\rm hot}$ (keV)   & 136 (f) \\
$\tau_{\rm hot}$ & 0.47$^{+0.04}_{-0.02}$\\
$\Gamma_{\rm hot}$ & 2.02$^{+0.01}_{-0.03}$ \\
$R_{\rm hot}$ (R$_{\rm g}$) & 17$\pm$1 \\
$kT_{\rm warm}$ (keV) & 0.32$^{+0.08}_{-0.04}$ \\
$\Gamma_{\rm warm}$ & 2.88$^{+0.06}_{-0.09}$  \\
$R_{\rm warm}$  (R$_{\rm g}$)& 43$^{+7}_{-9}$ \\
log\,$\xi$  &  0.5$\pm$0.2\\
A$_{\rm Fe}$  &  1.4$^{+0.4}_{-0.2}$\\
$norm_{\rm reflkerrd}$ ($\times$10$^{-3}$) & 4.3$\pm$0.6 \\
\hline
$F^{\rm unabs}_{\rm 0.001-100\,keV}$$^{(a)}$ (erg\,cm$^{-2}$\,s$^{-1}$) &   6.0$\times$10$^{-10}$ \\
$L^{\rm unabs}_{\rm 0.001-100\,keV}$$^{(a)}$ (erg\,s$^{-1}$)  &   2.0$\times$10$^{45}$ \\
$\chi^{2}$/d.o.f. & 1149.3/923  \\
\hline    \hline
\end{tabular}
\label{tab:SED}
\flushleft
\small{{\bf Notes}. $^{(a)}$ Source flux and luminosity corrected from Galactic absorption and reddening.}
\end{table}

\begin{figure*}[t!]
\begin{tabular}{ccc}
\includegraphics[width=0.92\columnwidth,angle=0]{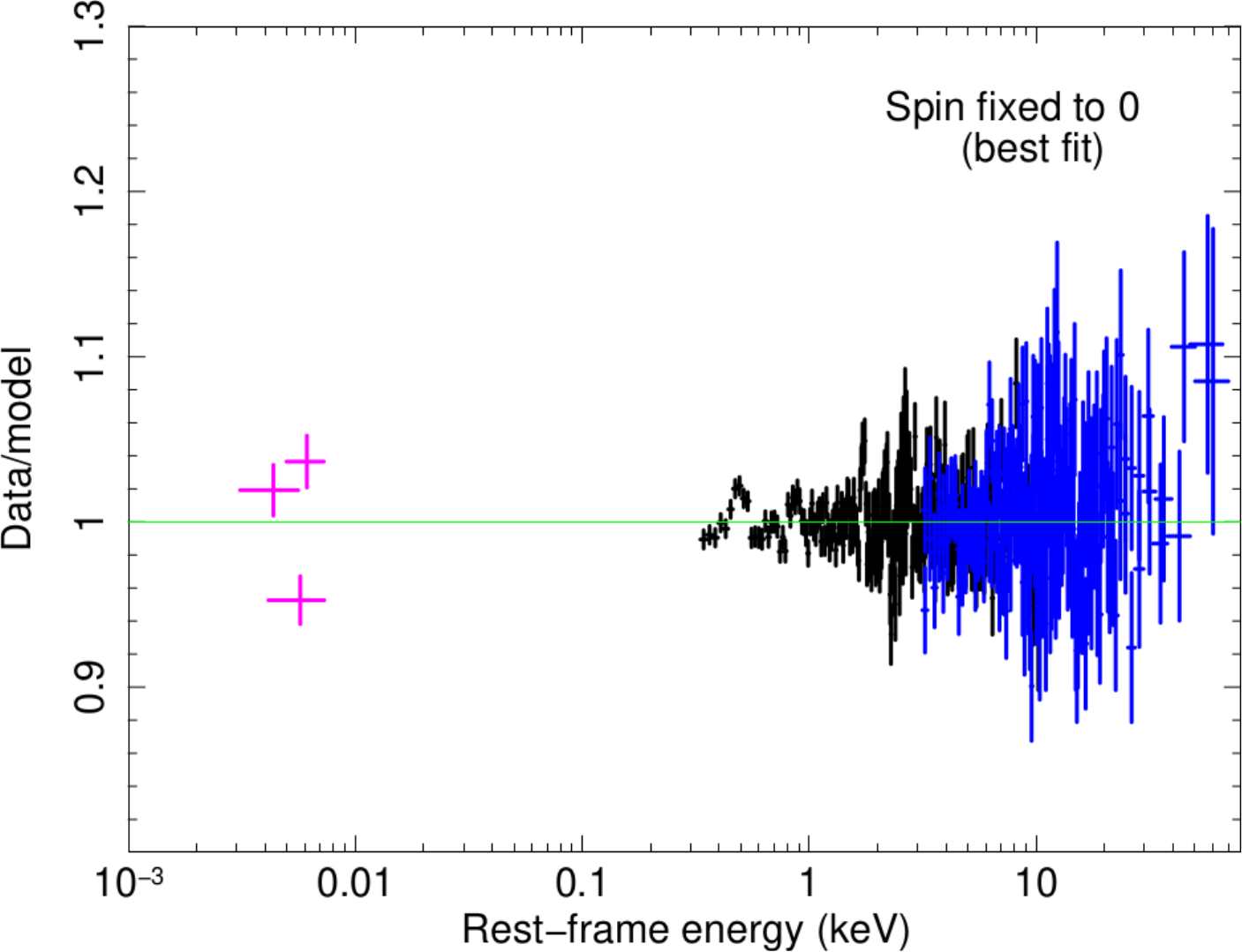} &
\includegraphics[width=0.92\columnwidth,angle=0]{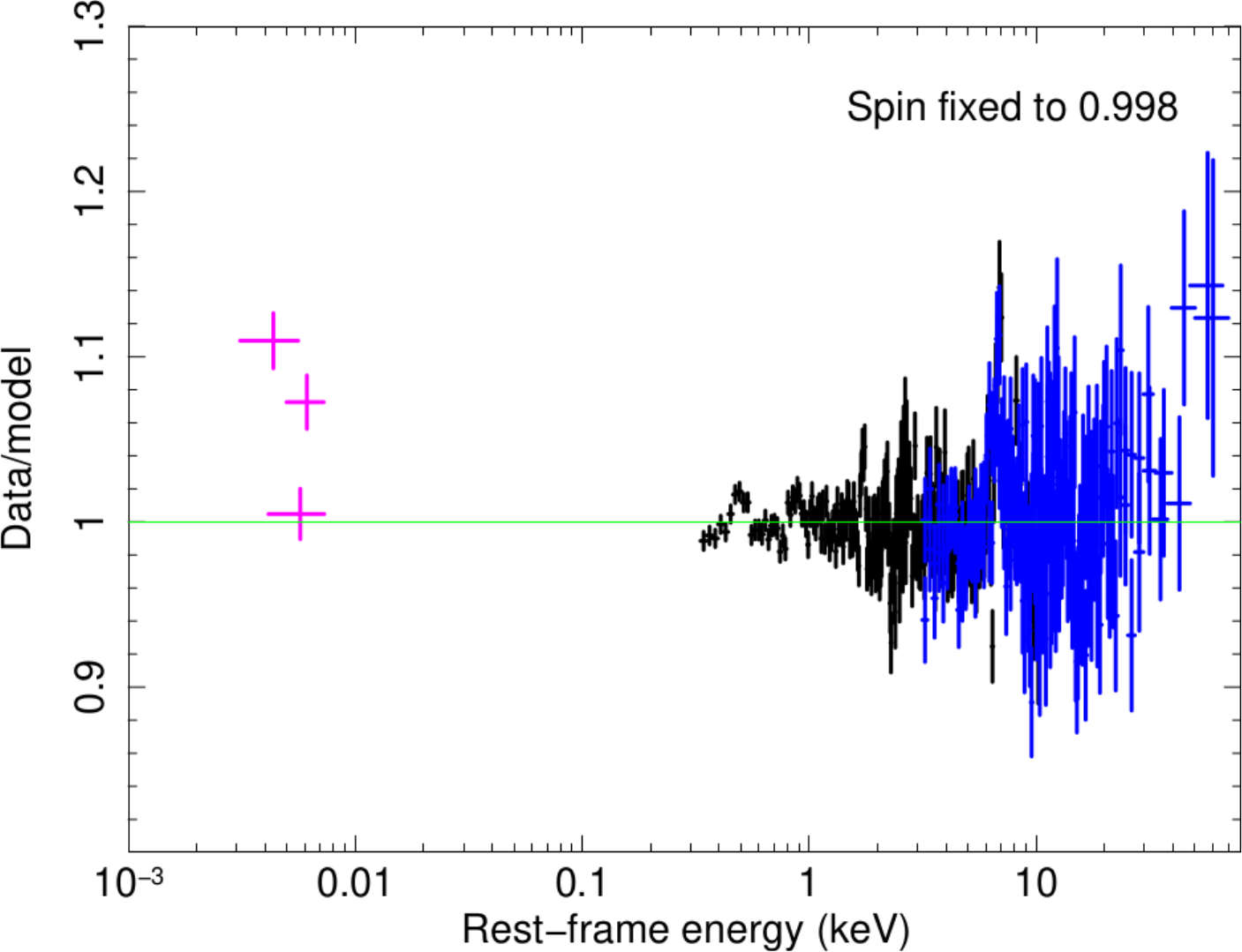}\\
\includegraphics[width=0.92\columnwidth,angle=0]{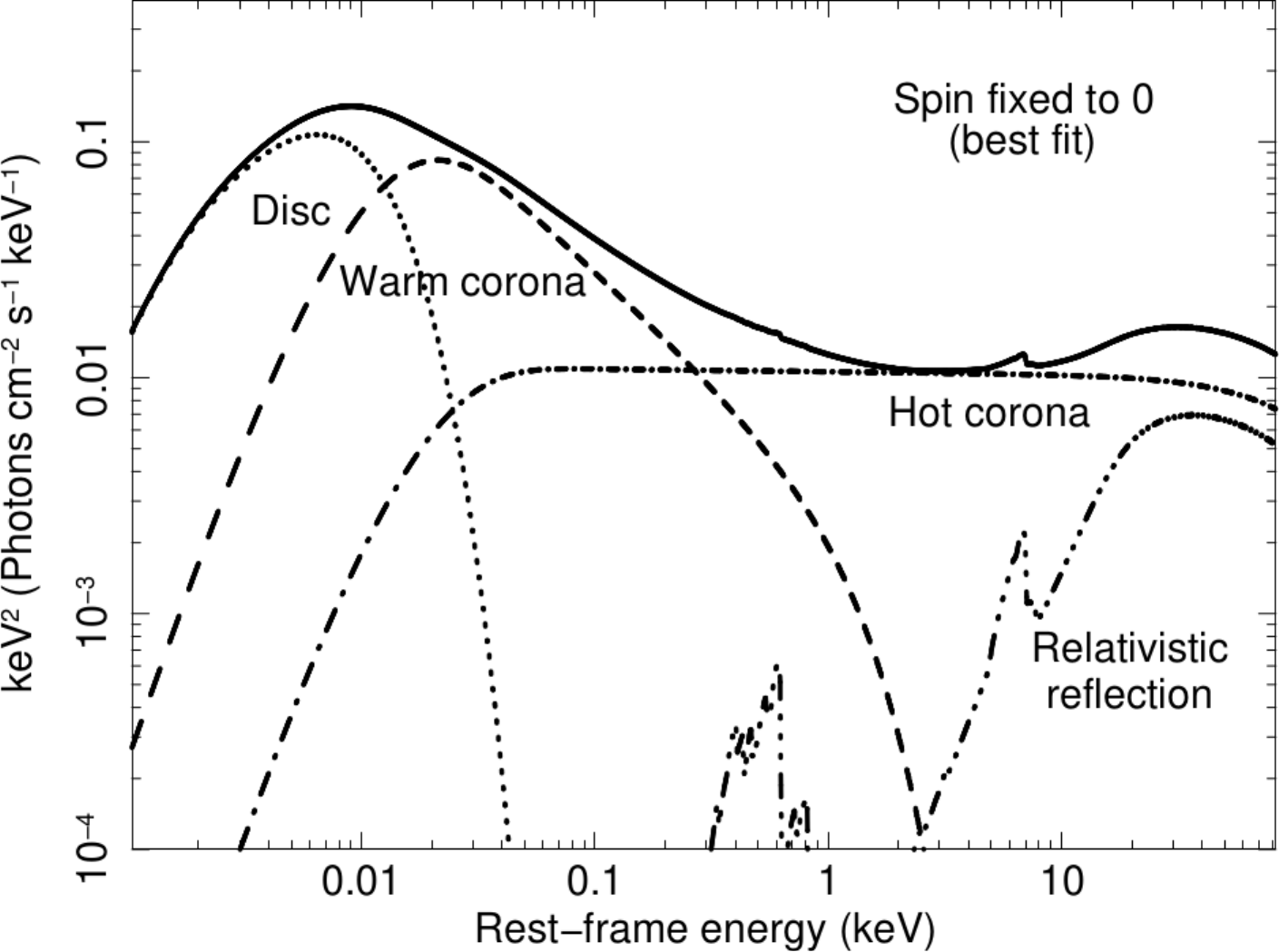} &
\includegraphics[width=0.92\columnwidth,angle=0]{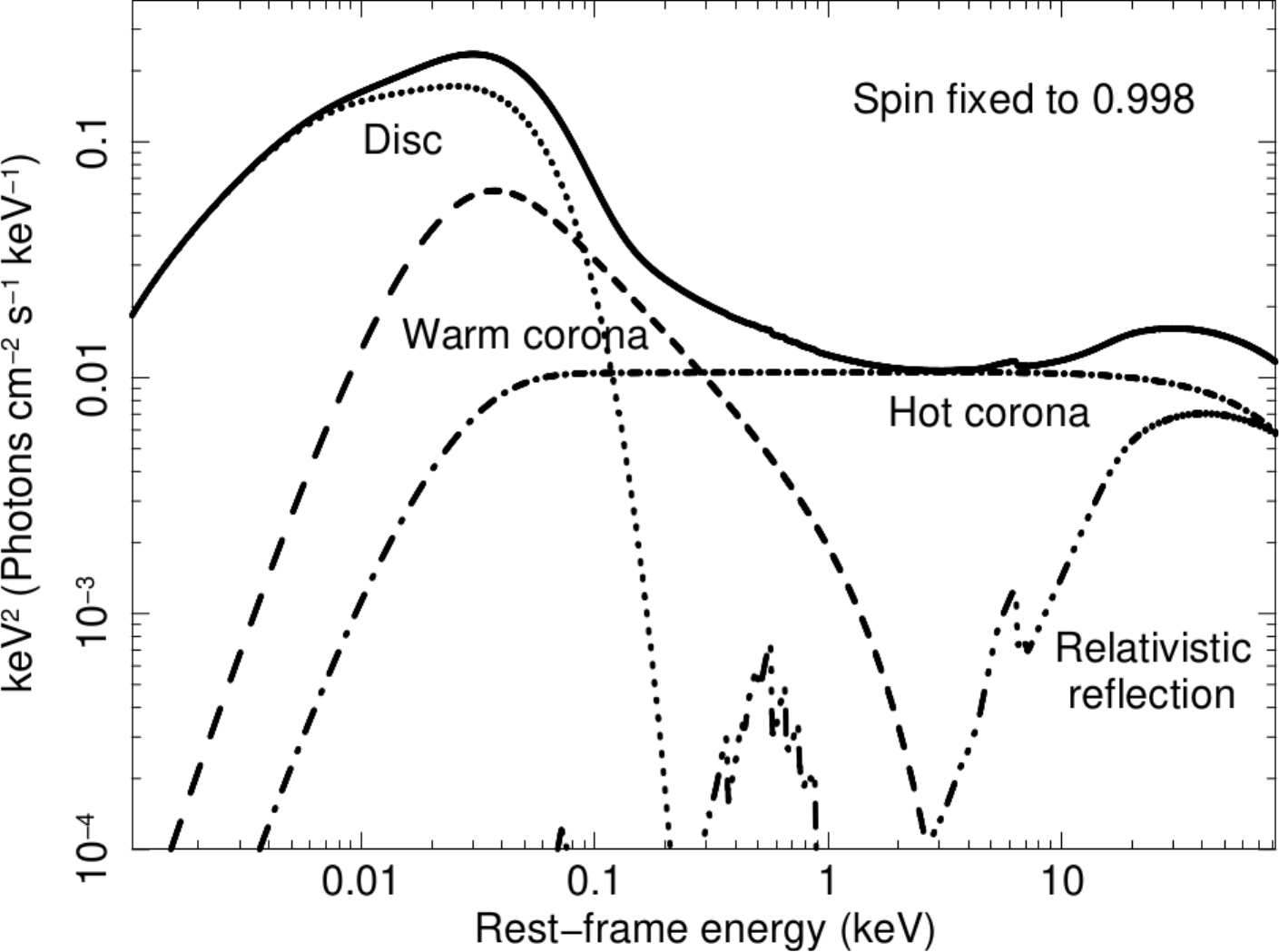} \\
\\
\end{tabular}
\caption{SED fits from UV to hard X-rays of ESO\,141-G55 using the {\sc relagn+reflkerrd} baseline model for the 2022 simultaneous {\sl XMM-Newton} (pn: black; OM: magenta) and {\sl NuSTAR} (blue) data, fixing the black hole spin to zero (left panels) and to 0.998 (right panels). The values of the best-fit parameters are reported in Table~\ref{tab:SEDspin}.  Top panels: data-to-model ratio. Bottom panels: intrinsic SED corrected for reddening and Galactic absorption (solid curves), with the main individual emission components: outer disc (dotted curves), warm optically-thick Comptonisation (dashed curves; warm corona), hot optically-thin Comptonisation (dotted-dashed curves; hot corona), and relativistic reflection (three-dotted-dashed curves). For clarity, the weak narrow Gaussian line component is not displayed.}
\label{fig:SED}
\end{figure*}

\begin{table*}[t!]
\caption{Simultaneous SED fits of the 2022 simultaneous {\sl XMM-Newton}-pn and {\sl NuSTAR} spectra of ESO141-G55  with the baseline model {\sc relagn+reflkerrd} for different fixed black hole spin values.}
\centering
\begin{tabular}{@{}l c c c c c c c}
\hline\hline
Parameters & $a$=0 & $a$=0.5 & $a$=0.7 &$a$=0.90 &$a$=0.95 & $a$=0.998 \\
\hline
log\,$\dot{m}$              &  $-$0.93$\pm$0.1 &  $-$0.85$^{+0.02}_{-0.01}$ &  $-$0.75$^{+0.02}_{-0.01}$   & $-$0.66$\pm$0.01 &$-$0.62$\pm$0.01 & $-$0.52$\pm$0.01\\
$\theta$ (degrees) & 43$^{+5}_{-4}$ & 41$^{+2}_{-1}$ &  41$\pm$1  & 34$\pm$1 & 25$\pm$1 & $\leq$9.2   \\
$kT_{\rm hot}$ (keV)   & 136(f)  & 136(f)  & 136(f) &  136(f)  & 136(f) & 136(f)\\
$\tau_{\rm hot}$ & 0.47$\pm$0.04 & 0.47$^{+0.05}_{-0.02}$  &  0.47$\pm$0.04 &0.53$^{+0.06}_{-0.08}$ & 0.59$^{+0.09}_{-0.08}$  & 0.73$^{+0.09}_{-0.08}$ \\
$\Gamma_{\rm hot}$ & 2.01$^{+0.01}_{-0.03}$ &  2.02$^{+0.02}_{-0.03}$ & 2.02$^{+0.02}_{-0.03}$  & 1.97$\pm$0.02 & 1.96$\pm$0.03  & 2.00$\pm$0.02\\
$R_{\rm hot}$ (R$_{\rm g}$) & 17$\pm$1 & 11$\pm$1 & 8.0$^{+0.3}_{-0.2}$& $\leq$6.2 & $\leq$6.1 & $\leq$6.0 \\
$kT_{\rm warm}$ (keV) & 0.33$^{+0.09}_{-0.06}$ & 0.32$^{+0.11}_{-0.06}$ & 0.33$^{+0.07}_{-0.06}$ & 0.55$^{+0.17}_{-0.12}$ & 0.55$^{+0.15}_{-0.09}$ & 0.44$^{+0.07}_{-0.04}$  \\
$\Gamma_{\rm warm}$ & 2.88$^{+0.06}_{-0.07}$ & 2.88$^{+0.06}_{-0.07}$ & 2.88$^{+0.06}_{-0.10}$ &2.93$\pm$0.05 & 2.92$^{+0.05}_{-0.06}$ & 2.96$^{+0.06}_{-0.07}$ \\
$R_{\rm warm}$  (R$_{\rm g}$)& 44$^{+5}_{-4}$ & 20$^{+4}_{-3}$ & 12$^{+3}_{-1}$ & 8.7$\pm$0.4 & 9.0$\pm$0.4 & 9.6$^{+0.6}_{-0.7}$  \\
log\,$\xi$ & 0.6$\pm$0.2 & 0.5$^{+0.3}_{-0.2}$ & 0.6$^{+0.3}_{-0.2}$ & 1.1$^{+0.2}_{-0.3}$ & 1.4$^{+0.3}_{-0.2}$ & 1.6$^{+0.2}_{-0.3}$   \\
A$_{\rm Fe}$  & 1.4$^{+0.4}_{-0.2}$ & 1.4$^{+0.4}_{-0.1}$& 1.5$^{+0.4}_{-0.2}$ & 2.2$^{+0.3}_{-0.4}$ & 1.9$^{+0.5}_{-0.3}$ & 0.9$\pm$0.1\\
$norm_{\rm reflkerrd}$ ($\times$10$^{-3}$) & 4.3$^{+0.7}_{-0.8}$ & 4.4$^{+0.7}_{-1.1}$ & 4.4$^{+0.7}_{-0.6}$ & 3.3$\pm$0.5  & 2.6$^{+0.6}_{-0.5}$ &  2.0$^{+0.4}_{-0.3}$ \\
\hline
$\chi^{2}$ (d.o.f.=924) & 1149.5 & 1157.6 & 1163.8 & 1187.1 & 1223.4 & 1319.1 \\ 
\hline    \hline
\end{tabular}
\label{tab:SEDspin}
\flushleft
\small{{\bf Notes}. `(f)' means that the value has been fixed.\\}
\end{table*}

Only the three shortest-wavelength UV filters with the OM (UVW2, UVM2, UVW1; effective wavelengths: 2120, 2310, 2910\,\AA, respectively) were used, as contamination by the host galaxy is negligible in these bands.
The Galactic reddening value, $E(B-V)$, of ESO\,141-G55 is 0.096 \citep{Schlafly11}.
For the colour correction of the outer disc, we apply the relation from \cite{Done12} by setting the $f_{\rm col}$  parameter model to a negative value.
The height of the vertical scale of the hot corona ($h_{\rm max}$ in $R_{\rm g}$) was tied to the radius of the hot corona ($R_{\rm hot}$). Since the SED fit is driven by the UV and soft X-ray data, we fixed the hot coronal temperature to 136\,keV, which is robustly inferred from the fit above 3\,keV (Sect.~\ref{sec:above3keV}). We set the distance of ESO\,141-G55 to 163.6\,Mpc \citep{Wright06,Planck20}, and its black hole mass to 1.3$\times$10$^{8}$\,M$_{\odot}$ \citep{Rokaki99}. The inclination of the accretion disc is free to vary.\\

A satisfactory fit from UV to hard X-rays is obtained with the values of the physical parameters reported in Table~\ref{tab:SED}, confirming a moderate Eddington accretion rate for ESO\,141-G55 of approximately 12\%. The warm corona temperature, $kT_{\rm warm}$=0.32$^{+0.08}_{-0.04}$\,keV,  is similar to that found using the simplified {\sc comptt} model above.  The intermediate value for the inclination angle of the disc, $\sim$43$^{\circ}$, is consistent with that found from the X-ray spectral analysis above 3\,keV ($\sim$44$^{\circ}$, Sect.~\ref{sec:above3keV}). The value for $R_{\rm hot}$ of 17\,$R_{\rm g}$ corresponding to the inner radius of the disc-corona system is in very good agreement with the inner radius where relativistic reflection occurs as inferred from the Fe\,K$\alpha$ profile (Sect.~\ref{sec:FeK}). The extension of the warm corona is about 2.5 times that of the hot corona.
From this best-fit model, an upper limit of 0.2 is found for the black hole spin, based on the classical $\Delta\chi^2$=2.706 criteria. \\

We check whether different values of the spin can be actually definitively excluded or can be considered as unlikely by fixing it to values of 0, 0.5, 0.7, 0.9, 0.95 and 0.998. 
As shown in Table~\ref{tab:SEDspin}, the value of $\chi^{2}$ (d.o.f.=924) increases with increasing spin value, by about $\Delta\chi^{2}$=+170 for a maximal spinning black hole. For a=0.9, $\Delta\chi^{2}$ is only +38, but the corresponding inclination of the disc is no longer consistent with the value inferred from the fit of the Fe\,K$\alpha$ relativistic line (Sect.~\ref{sec:FeK}) and from the fit above 3\,keV where the hot corona and the relativistic reflection dominate (Sect.~\ref{sec:above3keV}). This is even more pronounced at higher spin value where the inclination angle tends to the minimal value allowed by the {\sc reflkerrd} model, that is to say 9.1 degrees. Therefore, spin values above about 0.9 are found to be very unlikely for the disc-corona system of ESO\,141-G55. Figure~\ref{fig:SED} displays the data-model ratio of the fit for a non-spinning black hole (a=0, top left panel), and a maximal spinning black hole (a=0.998, top right panel), and illustrates how the UV data are badly fitted for a maximal black hole spin. This is illustrated in the lower panels of Fig.~\ref{fig:SED}, where the SED shapes corresponding to a non-spinning black hole (a=0, Fig.~\ref{fig:SED} is bottom-left) and a maximal spinning black hole (a=0.998 is bottom-right) are displayed.  For a=0.998, the peak emission of the outer disc (starting at $R_{\rm warm}$) shifts to higher energy (or in other words, has a higher temperature) compared to a=0, caused by the decrease of both the hot and warm corona radii, which is due to the energy balance used in {\sc relagn}. Indeed, as the black hole spin increases, $R_{\rm hot}$ must decrease in order to account for the higher emitting area (due to the decrease of the ISCO) and higher accretion efficiency. This in turn leads to a necessary decrease in $R_{\rm warm}$ \citep{Hagen23b}.

\section{Summary and discussion}\label{sec:discussion}

We presented the first ever broadband X-ray and SED analysis of the bright X-ray bare AGN ESO\,141-G55 based on the simultaneous observation with {\sl XMM-Newton} and {\sl NuSTAR} performed on October 1-2, 2022. This {\sl XMM-Newton} observation used the small window mode for the pn camera, which prevents pile-up for this bright X-ray source, compared to previous observations performed in 2001 and 2007, which suffer from very strong pile-up caused by the choice of the full frame mode. Furthermore, the 2022 {\sl XMM-Newton} and {\sl NuSTAR} observation ($\sim$124\,ks) provides the highest signal-to-noise ratio so far for ESO\,141-G55 from 0.3 to 79\,keV.
The simultaneous X-ray broadband spectrum of ESO\,141-G55 is characterised by the presence of a prominent (absorption-free) smooth soft X-ray excess, a broad Fe\,K$\alpha$ emission line and a significant Compton hump. 

\subsection{Presence of warm gas outside our line-of-sight}

The 2022 high-resolution {\sl XMM-Newton} RGS spectra confirmed the lack of intrinsic warm-absorbing gas associated with ESO\,141-G55, apart from the expected neutral absorption O\,\textsc{i} line associated with our Galaxy \citep{Gondoin03,Fang15,Gatuzz21}. 
Therefore, ESO\,141-G55 can still be classified as a bare AGN during this 2022 observation, with an upper limit of 4$\times$10$^{20}$\,cm$^{-2}$ for the column density of any warm absorbing gas in the AGN rest-frame in our line-of-sight. 
Nevertheless, we report the presence of several soft X-ray emission lines in the RGS spectrum, associated with the AGN and arising from the He- and H-like ions of oxygen and neon. 
In particular, the He-like resonance line profiles of \ion{O}{vii} and \ion{Ne}{ix} appear to be velocity broadened, with a typical FWHM of $\sim$3000--4000\,km\,s$^{-1}$ consistent with optical-UV BLR lines, while the profile of the forbidden \ion{O}{vii} line is unresolved and could be associated with a farther out region such as the NLR.
We note that the 2007\#4 RGS spectrum (the longest 2007 observation, Table~\ref{tab:log}) is similar to that of the 2022 one with the presence of a broad resonance line and a narrow forbidden line from \ion{O}{VII} (Fig.~\ref{fig:2007OVII}).
This adds evidence that bare AGNs are not intrinsically bare and that substantial X-ray-emitting gas is present out of our direct line-of-sight toward them, as also reported for the two prototype bare AGN Ark\,120 \citep{Reeves16} and Fairall\,9 \citep{Emmanoulopoulos11,Lohfink16}.
We note that the presence of the resonance line would suggest a significant contribution of photoexcitation \citep{Sako00,Kinkhabwala02}. 
 Enhancement of the intensity of the broad \ion{O}{vii} resonance line only (with no broad forbidden and/or intercombination lines observed) in a photoionised environment by continuum pumping would point out to an emitting region with a column density of $\sim$10$^{22}$--10$^{23}$\,cm$^{-2}$ and a high ionisation parameter $\xi$, on log scale, of $\sim$2--2.5 \citep{Chakraborty22}, as for example highly ionised gas located in the base of a radiatively driven disc wind at the outer BLR \citep{Risaliti11,Medhipour17}. An alternative origin from a collisional plasma cannot be definitively ruled out, such as signatures from shocked outflowing gas, hot circumnuclear gas, and/or star-forming activity \citep{Guainazzi09,Pounds11,Braito17,Grafton-Waters21,Buhariwalla23}. However, in the collisional plasma scenario, the lack of any corresponding (broad) forbidden and/or intercombination line would be difficult to explain, since the (forbidden+intercombination)/resonance line ratio is expected to be about unity \citep{Porquet01,Porquet10}.
The quality of the present RGS data prevent us from discriminating between photo-ionised and collisional plasma, and from probing resonance scattering, thanks to, for example, the detection of radiative recombination continua (RRC), Fe\,L (e.g., \ion{Fe}{xvii}) emission line ratio and/or high order ($n$>2) series emission lines \citep{Hatchett76,Liedahl90,Kallman96,Liedahl99,Sako00,Kinkhabwala02,Guainazzi07,Porquet10}. 
Only much higher signal-to-noise RGS spectra and/or observations with future soft X-ray spatial missions, such as the {\it Line Emission Mapper} (LEM) \citep{Kraft22}, {\sl Arcus} \citep{Smith16} and the {\sl Hot Universe Baryon Surveyor} \citep[HUBS;][]{Bregman23}, will allow us to determine the origin of the \ion{O}{vii} and \ion{Ne}{ix} broad resonance lines in ESO\,141-G55.

\subsection{A sporadic UFO in ESO\,141-G55?}

The intermediate inclination of the ESO 141-G55 disc-corona system, $\theta$$\sim$43$^{\circ}$, is a favourable configuration to intercept a possible disc wind, as suggested by the previous 2001 and 2007 XMM-Newton observations (Sect.~\ref{sec:UFO}). We then checked for a possible UFO signature in the present high signal-to noise 2022 observation, but we found no evidence for such a wind. 
In order to provide a direct and homogeneous comparison between the present 2022 and the 2007 dataset, we reprocessed and reanalysed the 2007 pn spectrum as detailed in Sec.~\ref{sec:UFO}. We confirm the presence, as first reported by \cite{DeMarco09},  of a narrow absorption feature during the October 30 2007 observation (2007\#4) at 8.34$\pm$0.05\,keV with an EW of 77$\pm$28\,eV. By modelling it with an {\sc xstar} absorber, we infer the following properties for the 2007\#4 UFO: N$_{\rm H}$=3.8$^{+3.5}_{-1.9}$$\times$10$^{23}$\,cm$^{-2}$, log\,$\xi$=4.5$^{+0.5}_{-0.2}$ and an outflow velocity $v_{\rm out}$=-0.176$\pm$0.006\,c.  These parameters are typical of other UFOs found in many AGNs \citep[e.g.,][]{Tombesi10,Gofford13,Matzeu23}.

In the case of ESO\,141-G55, the absorption arises purely from $\ion{Fe}{xxvi}$ and thus it originates from very highly photo-ionised gas. In comparison, adopting the same velocity and ionisation as above, the upper limit to the column density for the 2022 observation is $N_{\rm H}$$<$1.7$\times$10$^{22}$\,cm$^{-2}$, about an order of magnitude below what was found in the October 30 2007 observation. It is worth noting that for the combined pn spectrum of the three shorter {\sl XMM-Newton} observations performed between October 9 and October 12, 2007, that is to say only less than 3 weeks before the October 30 2007 observation, only an EW upper limit of 48\,eV is found for any UFO signature at the same energy. This could indicate a variable UFO on both short and long-time scales in ESO\,141-G55.  Variable UFO signatures have been observed in other AGNs, such as \object{PG 1211+143} \citep{Reeves18}, \object{PG\,1448+273} \citep{Reeves23} and \object{MCG-03-88-007} \citep{Braito22}. Such a sporadic feature in ESO\,141-G55, which has a moderate Eddington accretion rate, could be explained by a failed disc wind in the line-driven scenario \citep{Proga04,Proga05,Giustini19}, as observed, for example, in the lower flux states of \object{NGC\,2992} \citep{Marinucci20,Luminari23} when the wind is not able to launch.  The wind variability could also be due to the specific intermediate inclination angle of the system of ESO\,141-G55, $\sim$43$^{\circ}$, which may only intercept the edge of the wind near its opening angle, rather than viewing deeper into the wind. Then occasionally a denser wind clump could be present in our line-of-sight resulting in a sporadic  UFO signature, rather than a more persistent disc wind as seen in some of the  higher inclination systems above.\\

Intriguingly, the X-ray fluxes during these different {\sl XMM-Newton} observations were similar with only about 20\%  variability, making it somewhat unlikely that the change in the opacity of the absorption feature is due to a significant change in ionisation, as observed for example in \object{PDS\,456} during a major X-ray flare in 2018 \citep{Reeves21a}. Furthermore, the SED shapes from the UV to 10\,keV are comparable too, although the UV emission was slightly lower in 2007\#4 (Fig.~\ref{fig:SED2007}). As an illustration, we report the SED analysis of the October 30 2007 observation using the baseline {\sc relagn + reflkerrd} model in Appendix~\ref{sec:SED2007}, which displays very similar disc-corona properties compared to the 2022 one, and no drastic differences in the disc-corona properties are observed (Table~\ref{tab:SED2007}).
The appearance of this possible variable UFO signature in ESO\,141-G55 does not seem to be related to some specific properties of the disc-corona system. Only further monitoring with adequate sampling from day-to-month timescales of this source will allow us to understand the UFO duty cycle and origin.

\subsection{Relativistic reflection from the inner part of the accretion disc and weak contribution from the molecular torus}
The observed significantly broad Fe\,K$\alpha$ line (6.46$\pm$0.09\,keV) and the Compton hump suggest a significant contribution from reflection. The Fe\,K$\alpha$ line width (FWHM$\sim$50000\,km\,s$^{-1}$) points to an origin from the inner part of the accretion disc, $\sim$10\,$R_{\rm g}$, but not down to the ISCO. We find that the EW of the narrow Fe\,K$\alpha$ component ($\sim$40\,eV) is in the lower range of the values found for type-1 AGNs \citep[$\sim$30--200\,eV;][]{Liu10,Shu10,Fukazawa11,Ricci14}. This is consistent with the small torus covering factor of about 10\% or less for ESO\,141-G55 that was inferred by \cite{Esparza-Arredondo21} from an infrared and X-ray study. Such a weak Fe\,K$\alpha$ narrow component can be due to the `X-ray Baldwin' effect (also called the `Iwasawa-Taniguchi' effect), that is, to the anti-correlation between the EW of the narrow neutral core Fe K$\alpha$ line and the X-ray continuum luminosity related to the decrease of the covering factor of the molecular torus with the X-ray luminosity \citep[e.g.][]{Iwasawa93,Page04,Bianchi07,Shu10,Ricci14}. Indeed, according to the relationship between the EW of the Fe\,K$\alpha$ narrow core and the 2--10\,keV AGN luminosity obtained from a large AGN sample by \cite{Bianchi07}, an EW of about 50\,eV is expected for ESO\,141-G55 (L(2-10\,keV)=1.1$\times$10$^{44}$\,erg\,s$^{-1}$). This has been also observed in many bright individual X-ray AGNs with low or high Eddington accretion rates, such as Mrk\,110 \citep{Porquet19}, \object{IC\,4329A} \citep{Tortosa23}, PDS\,456 \citep{Reeves21a}, \object{I\,Zw\,1} \citep{Reeves19}, and PG\,1448+273 \citep{Reeves23}.

\subsection{The disc-corona system: presence of an optically-thin very hot corona and an optically-thick warm corona.}

The hot corona temperature ($kT_{\rm hot}$$\sim$120--140\,keV) does not vary significantly between the 2016 and 2022 {\sl NuSTAR} observations, where the flux increases only by about 10\%. The hot-corona temperature is found to be hotter than about 80\% of known Seyfert\,1s \citep{Lubinski16,Middei19,Akylas21}. 
An intermediate value for the inclination angle of the disc-corona system is found, $\sim$43$^{\circ}$ using complementary fitting methods: Fe\,K line profile (Sect.~\ref{sec:FeK}), relativistic reflection emission above 3\,keV (Sect.~\ref{sec:above3keV}) and SED (Sect.~\ref{sec:SED2022}). Moreover it is similar to the value inferred from accretion disc fits to the UV continuum and the H$\beta$ emission line, $\sim$42 degrees for a non-spinning black hole \citep{Rokaki99}, all indicating a similar inclination at different spatial scales of the accretion disc, thus no significant warp. \\

We found that relativistic reflection alone, albeit significant, cannot reproduce the simultaneous 2022 X-ray broadband spectra of ESO\,141-G55 as also found for the bare AGN Ark\,120 \citep{Matt14,Porquet18,Porquet19}, TON\,S180 \citep{Matzeu20} and Mrk\,110 \citep{Porquet24}. Indeed to reproduce the soft X-ray excess shape a much steeper and high relativistic reflection fraction are required compared to the hard X-ray energy range.
This reinforces once again the importance of considering the simultaneous X-ray broadband spectra up to 79\,keV when determining the processes at work in the disc-corona system, especially for the origin of the soft X-ray excess \citep{Matt14,Porquet18,Matzeu20,Porquet21}.\\

The SED (from UV to X-rays) is nicely reproduced by a model which combines the emission from a warm and hot corona, and outer disc ({\sc relagn}; \citealt{Hagen23b}) and relativistic reflection ({\sc reflkerrd}; \citealt{Niedzwiecki19}). 
The best fit points to an upper limit of 0.2 for the black hole spin (at the 90\% confidence level). The SED fit becomes significantly worse, particularly in the UV domain, as the spin increases toward its maximum value. Furthermore, the inferred values of the inclination angles of the disc are no more compatible for a$\geq$0.9 with that found from the fit above 3\,keV (see above). The Eddington accretion rate is found to be moderate at about 10--20\% (depending on the black hole spin). The ratio of the UV and X-ray peak emission ($\sim$ factor of 10) for ESO\,141-G55 is much higher than the one found for Mrk\,110 \citep{Porquet24}, but comparable to that of Fairall\,9 \citep{Hagen23a} and Ark\,120 \citep{Porquet19}.\\ 

The temperature (kT$_{\rm warm}$$\sim$0.34\,keV) and optical depth ($\tau_{\rm warm}$$\sim$12) of the warm corona inferred for ESO\,141-G55 are similar to those found for other low to moderate-Eddington accretion AGNs, i.e. kT$_{\rm warm}$$\sim$0.2--1\,keV and $\tau_{\rm warm}$$\sim$10--20 
\cite[e.g.][]{Porquet04a,Bianchi09,Petrucci13,Mehdipour15,Matt14,Porquet18,Middei19,Porquet19,Ursini20,Porquet24}. The physical existence of such an optically-thick warm corona is strengthened by several recent theoretical models of the disc-corona structure \citep{Petrucci18,Ballantyne20a,Ballantyne20b,Gronkiewicz23,Kawanaka24}. Importantly, the warm corona scenario can not only (mainly or totally) account for the soft X-ray excess but also for a significant contribution to the observed UV emission (in addition to the thermal disc emission) of ESO\,141-G55 (Fig.~\ref{fig:SED}, lower panels). This confirms previous studies reported for the {\it SOUX} AGN sample \citep{Mitchell23} and individual AGNs (e.g., \object{Mrk\,590}: \citealt{Mehdipour11}; \object{Ark\,120}: \citealt{Porquet19}; \object{Fairall\,9}: \citealt{Hagen23a};  and \object{Mrk\,110}: \citealt{Porquet24}). Moreover, it can naturally explain the simultaneous significant dimming of both the UV and soft X-ray excess emission observed in some AGNs, such as \object{ESO 511-G030} \citep{Middei23}, and \object{Mrk\,841} \citep{Mehdipour23}. The optically-thick warm corona scenario seriously challenges the standard disc theory in AGNs.

\begin{acknowledgements}
The authors thank the anonymous referee for useful and constructive comments.
The paper is based on observations obtained with the {\sl XMM-Newton}, an ESA science
mission with instruments and contributions directly funded by ESA
member states and the USA (NASA). 
 This work made use of data from the
{\sl NuSTAR} mission, a project led by the California Institute of
Technology, managed by the Jet Propulsion Laboratory, and
funded by NASA. 
This research has made use of the {\sl NuSTAR} 
Data Analysis Software (NuSTARDAS) jointly developed by
the ASI Science Data Center and the California Institute of
Technology.  This research has made use of the SIMBAD
database, operated at CDS, Strasbourg, France. This research has made use of the
NASA/IPAC Extragalactic Database (NED) which is operated by the California
Institute of Technology, under contract with the National Aeronautics and Space Administration. This work was supported by the French space agency (CNES). This research has made use of the computing facilities operated by CeSAM data centre at LAM, Marseille, France. SH acknowledges support from the Science and Technology Facilities Council (STFC) through the studentship ST/V506643/1.
\end{acknowledgements}

\bibliographystyle{aa}
\bibliography{biblio}

\clearpage

\appendix

%---------------------------------
\section{2007 {\sl XMM-Newton} observations}\label{sec:2007obs}

Here, we present the analysis of the four 2007 {\sl XMM-Newton} observations (see Table~\ref{tab:log}), which are relevant for our comparison with the 2022 observation. The extraction of all 2007 pn light curves (Sect.~\ref{sec:2007lc}) and of the 2007\#4 pn spectra have been built using an annulus extraction region to minimise the effect of heavy pile-up due to the use of the Full Frame Window mode (see Sect.~\ref{sec:xmm}). The reprocessing and extraction method for the 2007\#4 RGS spectra are reported as well in Sect.~\ref{sec:xmm}.  

\subsection{2007 {\sl XMM-Newton}-pn light curves}\label{sec:2007lc}

\begin{figure}[t!]
\begin{tabular}{c}
\includegraphics[width=0.9\columnwidth,angle=0]{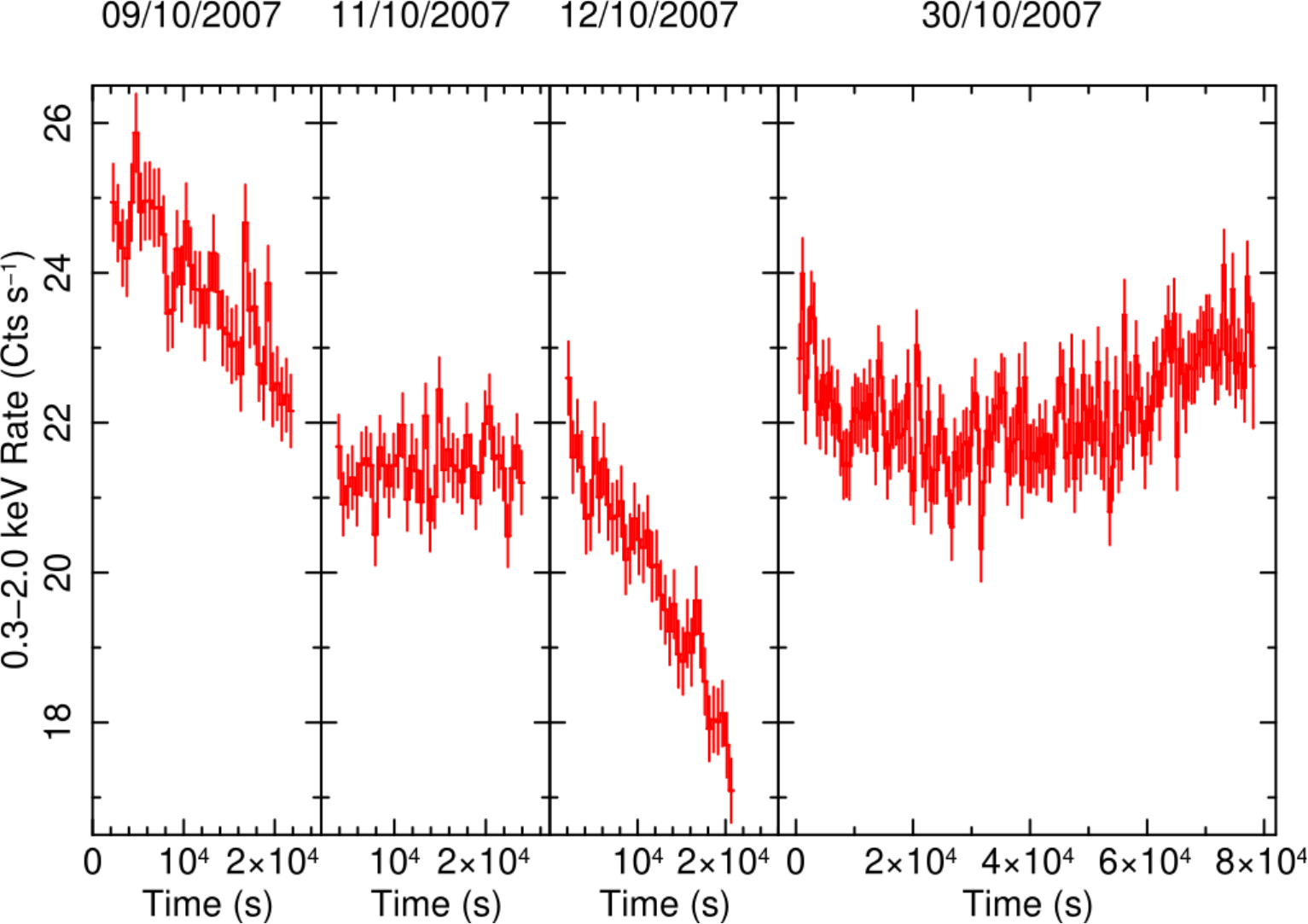} \\
\includegraphics[width=0.9\columnwidth,angle=0]{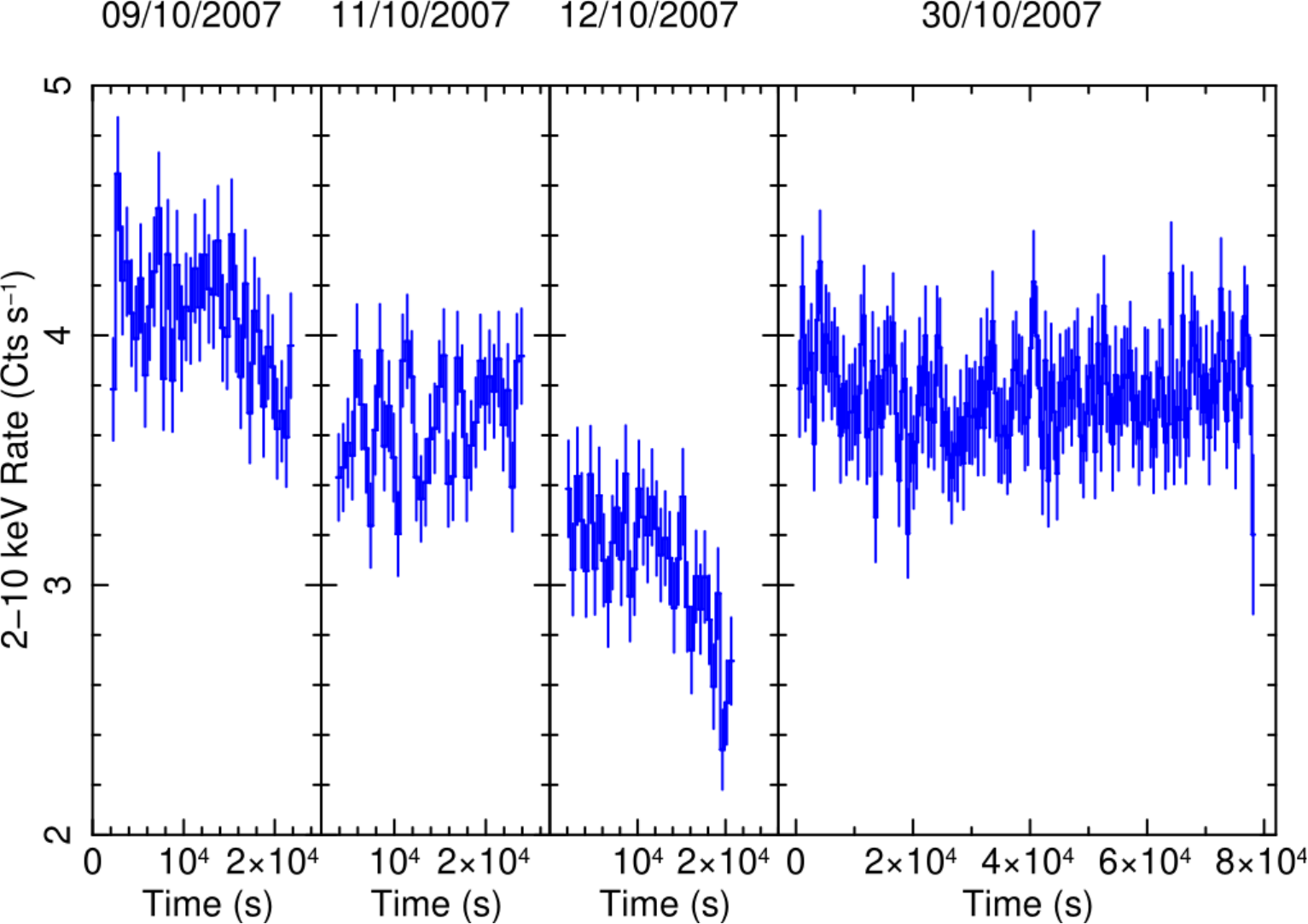} \\
\includegraphics[width=0.9\columnwidth,angle=0]{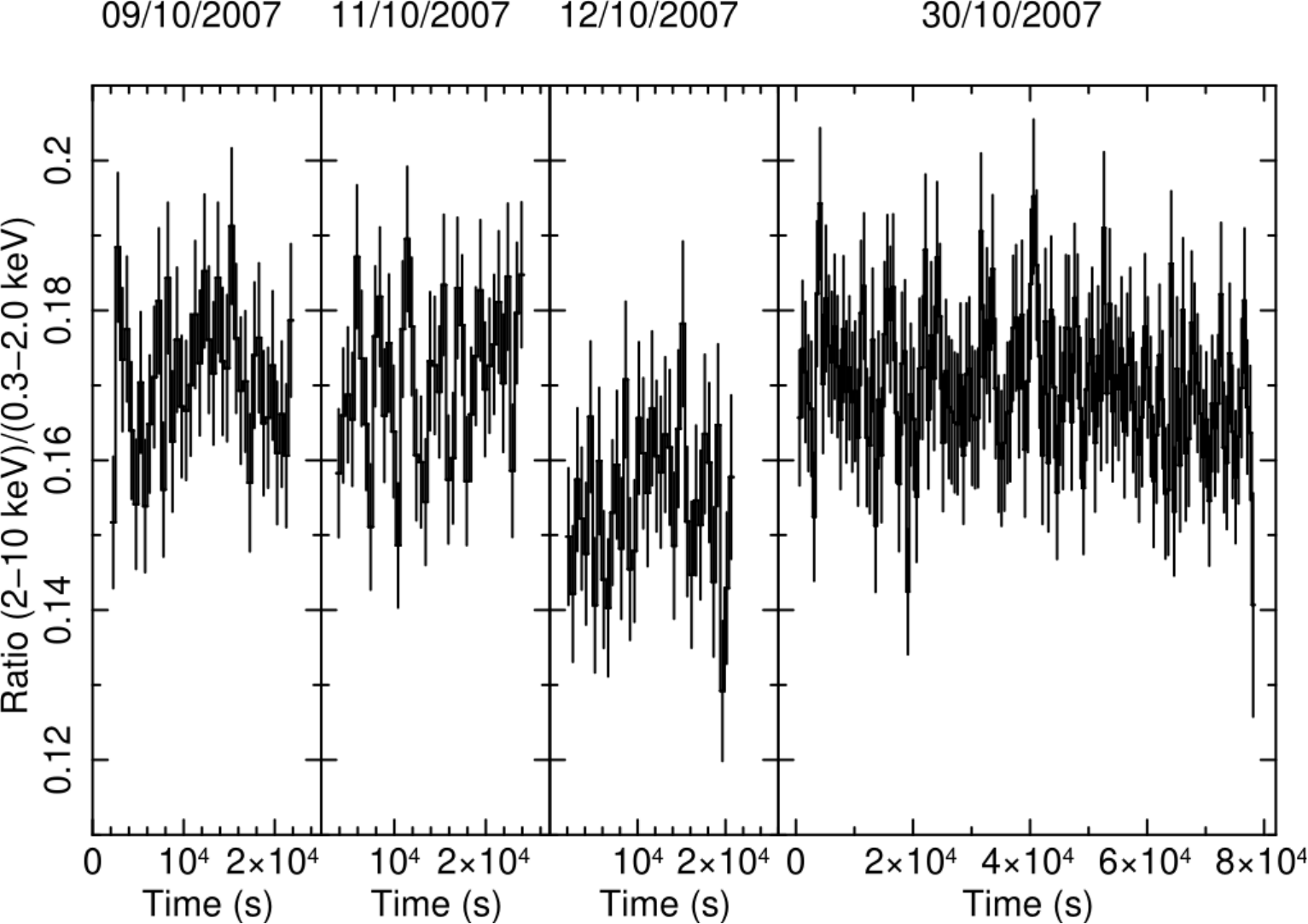} \\
\\
\end{tabular}
\caption{The 2007 {\sl XMM-Newton}-pn light curves of ESO\,141--G55, with time bins of 500 s, in the 0.3--2\,keV (top panel) and 2-10\,keV (middle panel) energy ranges and the corresponding hardness ratio, 2--10\,keV/0.3--2\,keV (bottom panel).} 
\label{fig:2007lc}
\end{figure}

The count rate light curves of the four {\sl XMM-Newton}-pn observations have been built using the {\sc epiclccorr} {\sc SAS} tool. Figure~\ref{fig:2007lc} shows the corresponding light curves in the 0.3--2\,keV and 2--10\,keV energy bands, and the hardness ratio, 2-10\,keV/0.3--2\,keV. While the observation performed on 09/10/2007 and 12/10/12/2007 exhibit some variability, mainly in the soft X-rays similar to the 2022 observation, no significant spectral variability is reported from the hardness ratio at each epoch.\\

\subsection{RGS spectrum of the 2007\#4 observation}\label{sec:2007RGS}

\begin{figure}[t!]
\begin{tabular}{c}
\includegraphics[width=0.95\columnwidth,angle=0]{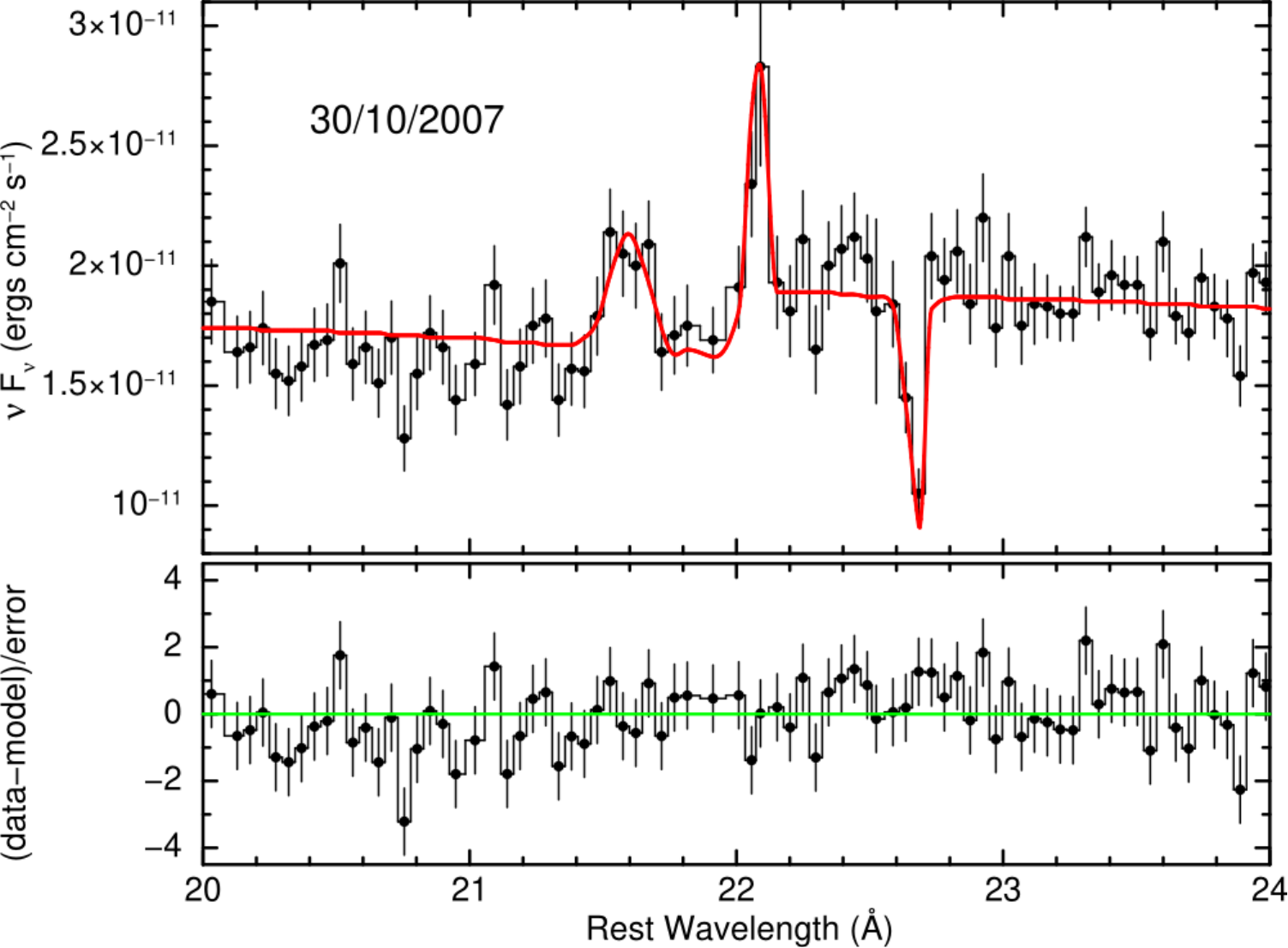} \\
\end{tabular}
\caption{A zoom in the \ion{O}{vii} emission band of the fluxed 2007\#4 RGS spectrum of ESO\,141-G55 fitted with a baseline absorbed power-law continuum. Similar to the 2022 RGS spectrum (Fig.~\ref{fig:RGS} bottom panel) the broad component at 21.6\,\AA\ corresponds to the resonance line and the narrow component at 22.1\,\AA\ corresponds to the forbidden line. The absorption line at about 22.6\,\AA\ is due to \ion{O}{i} from our Galaxy.}
\label{fig:2007OVII}
\end{figure}

Figure~\ref{fig:2007OVII} displays a zoom-in on the \ion{O}{vii} triplet obtained from the 2007\#4 (longest 2007 observation) RGS spectrum using an absorbed power-law model for the continuum as done for the 2022 RGS spectrum (Sect.~\ref{sec:rgs}). 
Similar to the observation in 2022, a broad resonance line and a narrow forbidden line are present. The parameters are also consistent with the 2022 observation; here the resonance line is centered at 
$E=574.0\pm1.7$\,eV, with an equivalent width of $1.4\pm0.5$\,eV, 
while the line is somewhat broadened with a FWHM velocity width of about 3000\,km\,s$^{-1}$. The forbidden line is centered at $E=561.8\pm1.7$\,eV, with an equivalent width of $1.0\pm0.6$\,eV and is unresolved.

%---------------
\section{Hot corona and relativistic properties inferred from the relxill models: 2022 and 2016 epochs}\label{app:relxill}

\begin{figure}[t!]
\begin{tabular}{c}
\includegraphics[width=0.9\columnwidth,angle=0]{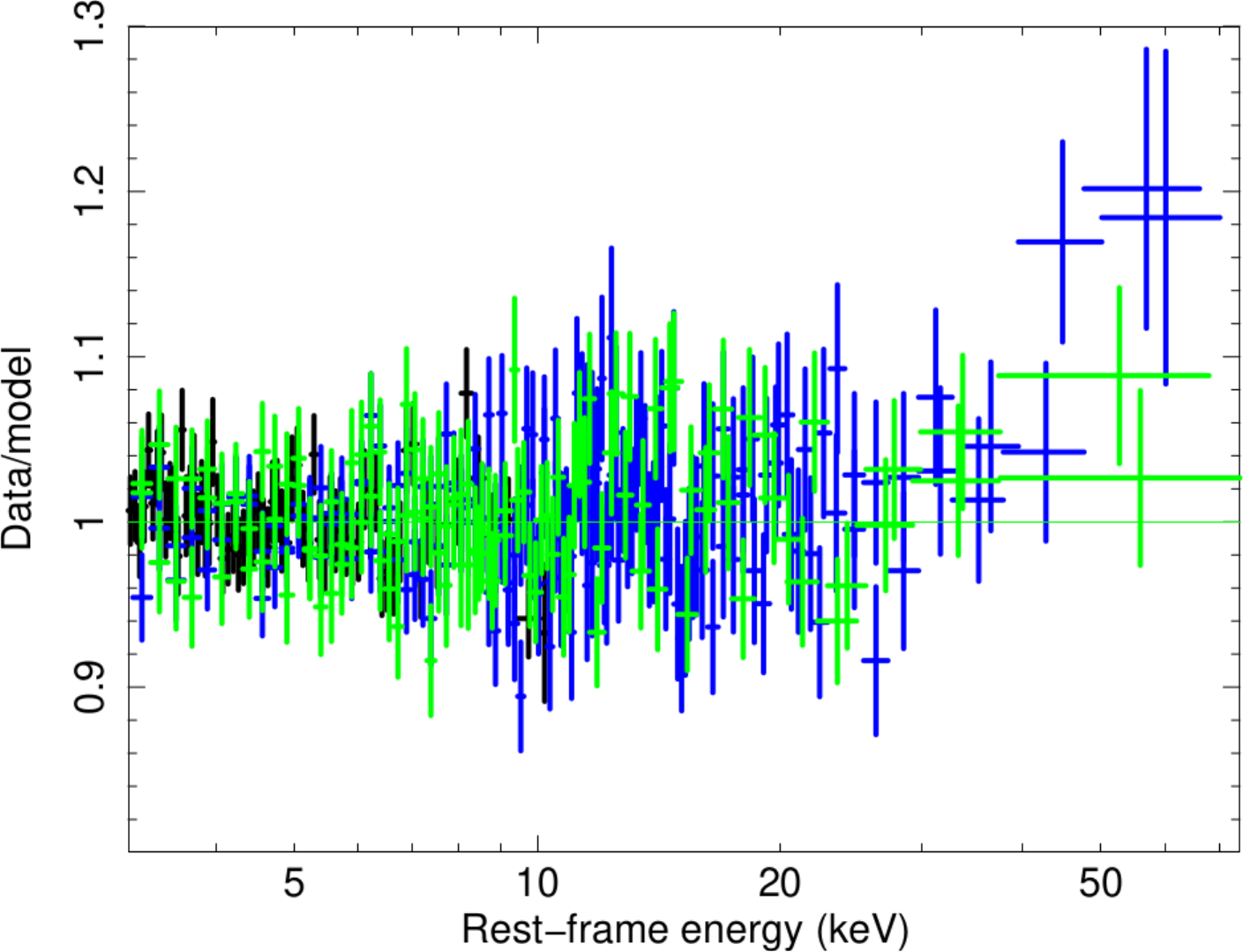}
 \end{tabular}
\caption{Data-to-model ratio of the fit above 3\,keV of the 2022 simultaneous {\sl XMM-Newton} (black) and {\sl NuSTAR} (blue) spectrum and of the 2016 {\sl NuSTAR} (green) spectrum using the {\sc relxillcp} relativistic reflection model with a primary Comptonisation continuum shape.}
\label{fig:relxillcpabove3keV}
\end{figure}

\begin{figure}[t!]
\begin{tabular}{c}
\includegraphics[width=0.9\columnwidth,angle=0]{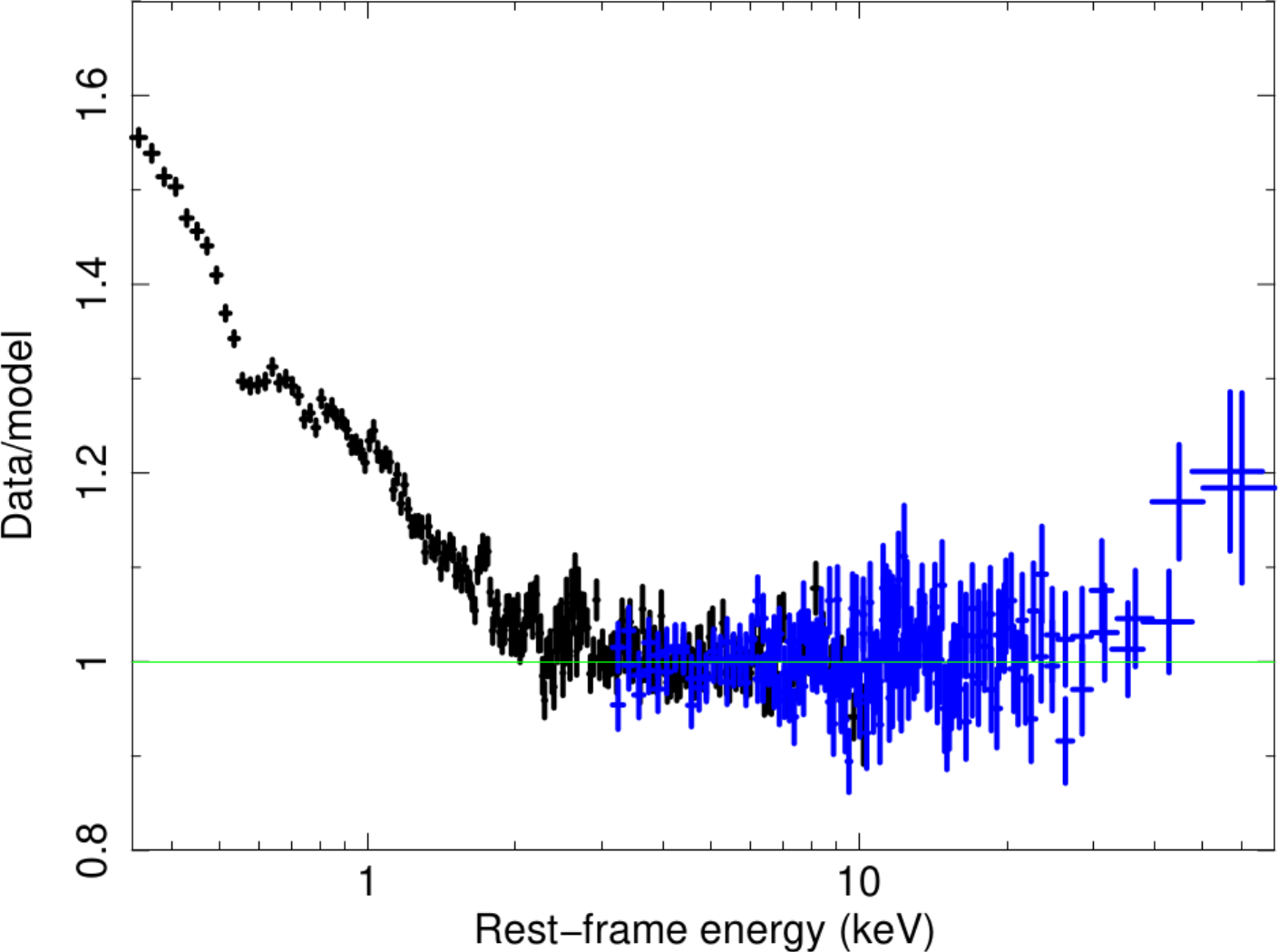} \\
\includegraphics[width=0.9\columnwidth,angle=0]{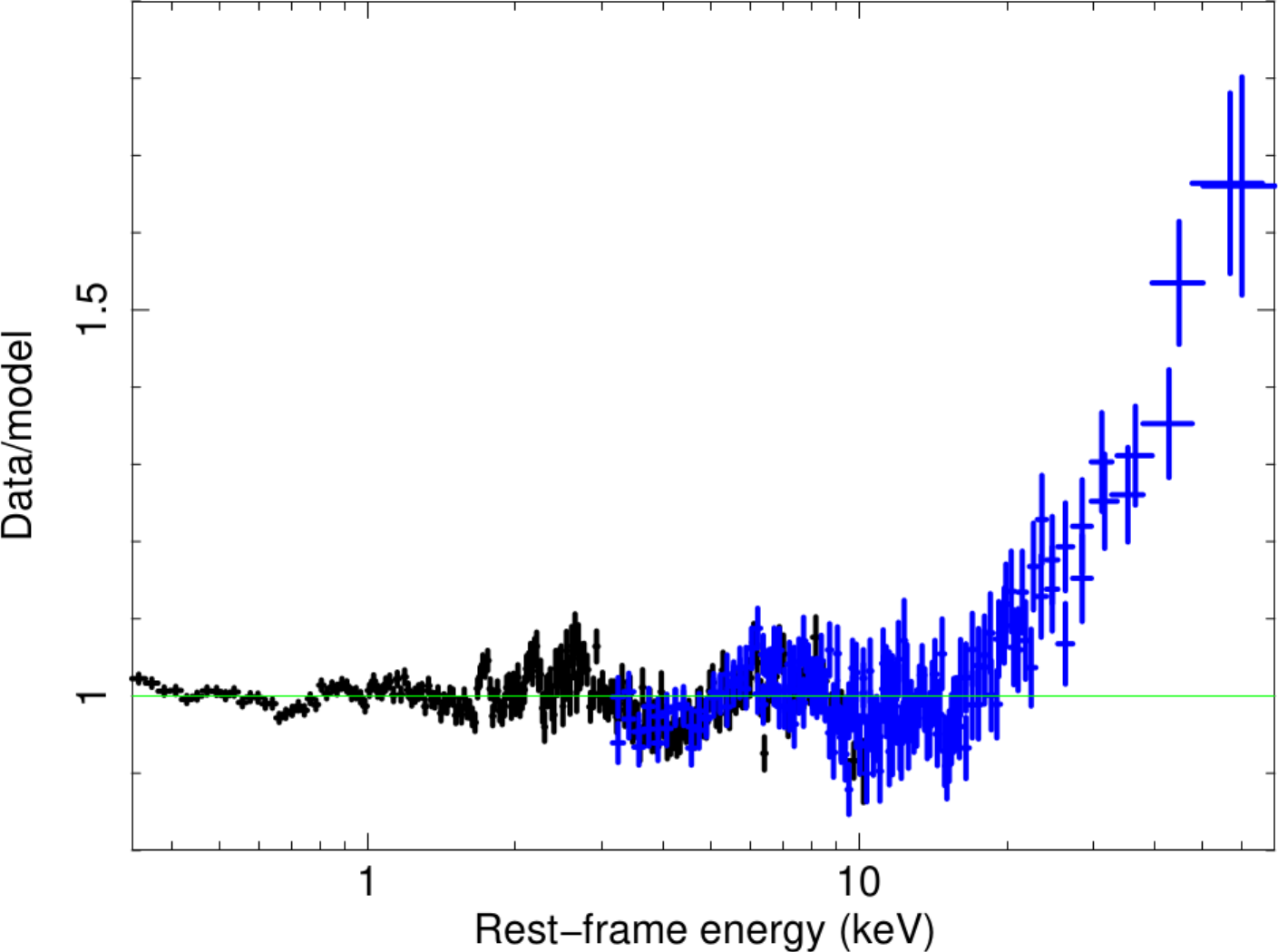}\\
\end{tabular}
\caption{Residuals in term of data-to-model ratio for the fits of the 2022 simultaneous {\sl XMM-Newton}-pn (black) and {\sl NuSTAR} (blue) spectrum of ESO\,141-G55 with the {\sc relxillcp} model.
	Top panel: fit above 3\,keV (Table~\ref{tab:relxillcpabove3keV}), which has been extrapolated down to 0.3\,keV.
       Bottom panel: fit over the full 0.3--79\,keV energy range.}
\label{fig:relxillcp}
\end{figure}

We report the pn and {\sl NuSTAR} spectral analysis above 3\,keV using the {\sc relxillcp} model \citep{Dauser13}.
As mentioned in Sect.~\ref{sec:above3keV}, the {\sc relxillcp} model uses the {\sc nthcomp} model \citep{Zdziarski96,Zycki99} to reproduce the hot corona (hard Comptonisation) spectral shape, while for the {\sc reflkerrd} model (Sect.~\ref{sec:above3keV}), the hard Comptonisation is calculated using {\sc ireflect} convolved with {\sc compps} \citep{Poutanen96}. 
As done in Sect.~\ref{sec:above3keV}, the single power-law disc emissivity index ($q$; with emissivity $\propto R^{-q}$) is allowed to vary, while the inner disc radius was set to the ISCO. Since the spin value is not constrained, we fixed it at zero.
The best-fit parameters are reported in Table~\ref{tab:relxillcpabove3keV}.
Even if the fit is statistically satisfactory, a noticeable positive residual is present above $\sim$40\,keV for the 2022 observation (Fig.~\ref{fig:relxillcpabove3keV}) and only lower limits for the hot corona temperature are found at both epochs. This could be due to the use of {\sc nthcomp} model to reproduce the hard Comptonisation shape.
As for the {\sc reflkerrd} model, the extrapolation of the {\sc relxillcp} fit down to 0.3\,keV shows that the soft X-ray excess is not accounted for, leaving a huge positive residual below 2\,keV (Fig.~\ref{fig:relxillcp}, top panel).
When this model is applied to the whole 0.3--79\,keV X-ray broadband range, it fails to reproduce both the soft and hard X-ray shapes (Fig.~\ref{fig:relxillcp}, lower panel).
Similar results are found if a lamppost geometry is assumed for the hot corona using the {\sc relxilllpcp} model. 

\begin{table}[t!]
 \caption{Simultaneous fits above 3\,keV, using the {\sc relxillcp} relativistic reflection model, of the October 2022 {\sl XMM-Newton+NuSTAR} and July 2016 {\sl NuSTAR} observations of ESO\,141-G55.}
\centering
\begin{tabular}{@{}l l l }
\hline\hline
 Parameters               &   \multicolumn{1}{c}{2022}   &   \multicolumn{1}{c}{2016} \\
                          &   \multicolumn{1}{c}{\sl NuSTAR} &   \multicolumn{1}{c}{\sl NuSTAR}\\
                          &   \multicolumn{1}{c}{+{\sl XMM}} & \\
\hline
\hline
$\theta$ (deg) & \multicolumn{2}{c}{44$\pm$3} \\
$kT_{\rm hot}$ (keV)                 &  $\geq$223               & $\geq$82         \\
$\Gamma_{\rm hot}$                     & 2.01$^{+0.03}_{-0.02}$           & 1.94$\pm$0.03     \\
$q$           & \multicolumn{2}{c}{2.0$^{+0.4}_{-0.3}$}                  \\
log\,$\xi$ (erg\,cm\,s$^{-1}$)       &   \multicolumn{2}{c}{$\leq$1.6}     \\
$A_{\rm Fe}$ & \multicolumn{2}{c}{1.2$^{+0.4}_{-0.2}$} \\
$\cal{R}$                                  & 0.89$^{+0.16}_{-0.13}$  &  0.59$\pm$0.12        \\
$norm_{\rm relxillcp}$ ($\times$10$^{-4}$) & 1.5$\pm$0.1        & 1.6$\pm$0.1     \\
$\chi^{2}$/d.o.f.  ($\chi^2_{\rm red}$) & \multicolumn{2}{c}{1270.2/1194 (1.06)}  \\
\hline
\hline
\end{tabular}
 \label{tab:relxillcpabove3keV}
\end{table}

%--------------------

\section{SED analysis of the October 30, 2007 {\sl XMM-Newton} observation}\label{sec:SED2007}

\begin{figure}[t!]
\begin{tabular}{c}
\includegraphics[width=0.95\columnwidth,angle=0]{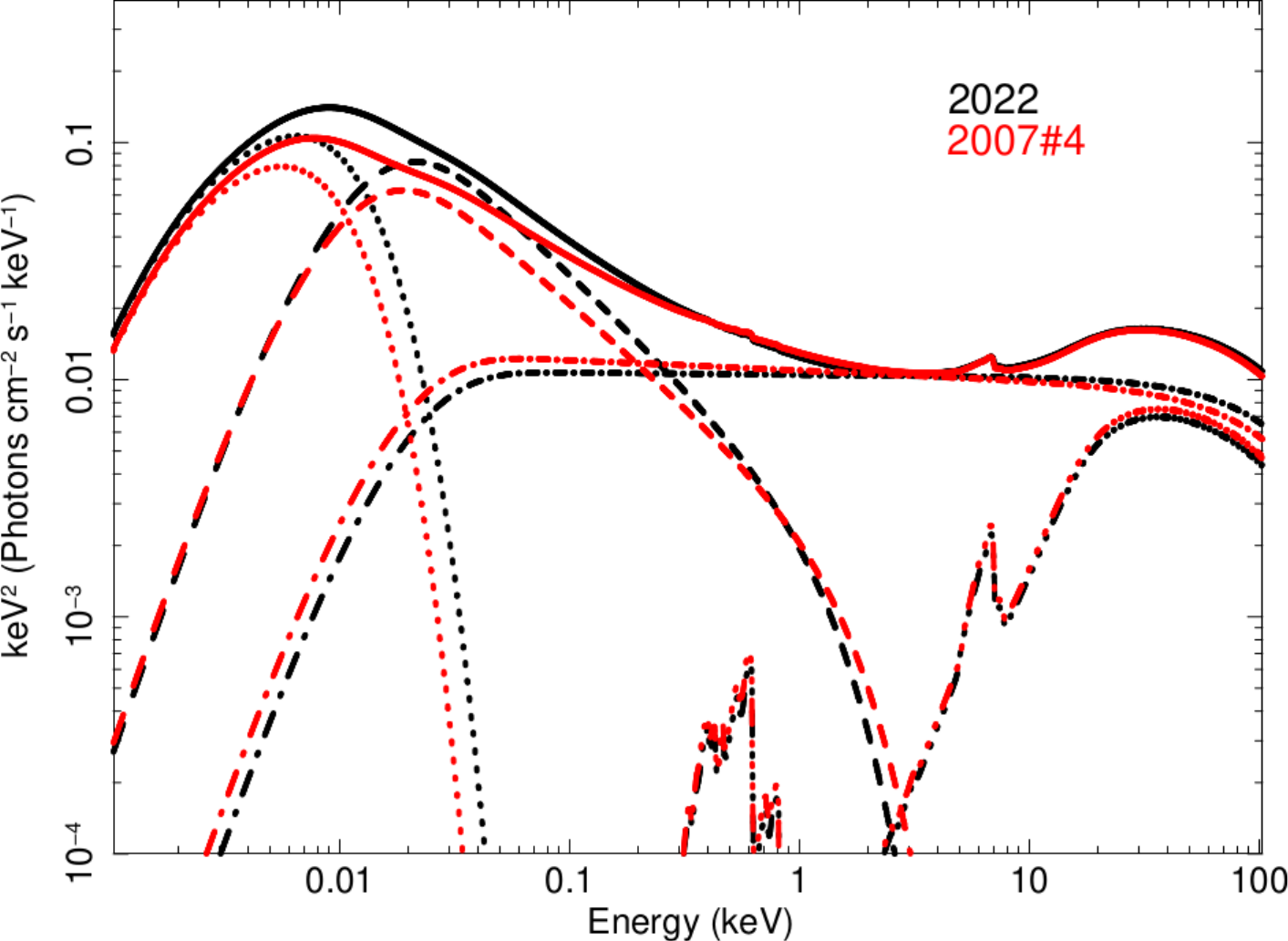} \\
\\
\end{tabular}
\caption{Intrinsic SED of ESO\,141-G55 inferred from the {\sc relagn+reflkerrd} baseline model for the 2007\#4 {\sl XMM-Newton} observation (red). The best-fit parameter values are reported in Table~\ref{tab:SED2007}. For comparison, the 2022 intrinsic SED is also displayed (black). These intrinsic SEDs are corrected for both reddening and Galactic absorption (solid curves). The main individual emission components are displayed: outer disc (dotted curves), warm optically-thick Comptonisation (dashed curves; warm corona), hot optically-thin Comptonisation (dotted-dashed curves; hot corona), and relativistic reflection (three-dotted-dashed curves).}
\label{fig:SED2007}
\end{figure}

We present the SED analysis of the 2007\#4 {\sl XMM-Newton} (OM-UV plus pn) observation using the baseline model {\sc relagn+reflkerrd} (Sect.~\ref{sec:SED2022}). Our aim is to check for any significant difference of the properties of the disc-corona system, compared to those inferred from the 2022 observation, which could explain the presence of the  narrow UFO signature reported during the 2007\#4 observation (Sect.~\ref{sec:UFO}). The black hole spin value was fixed to zero. We set the values of the disc inclination and the iron abundance to those inferred from the 2022 SED analysis (see Table~\ref{tab:SEDspin}), since they are not supposed to vary on yearly timescales. Due to the lack of 2007 data above 10\,keV, we fixed the hot corona temperature to 136\,keV as found in 2022, since, as shown in Sect.~\ref{sec:above3keV} this parameter does not show any significant change for 10--20\% X-ray flux variation.  The best-fit parameters are reported in Table~\ref{tab:SED2007} and shows that the disc-corona properties are very comparable at both epochs taking into account error bars of the parameter values. Figure~\ref{fig:SED2007} shows, for illustration purposes, a comparison of the intrinsic SED at both epochs. Therefore, the appearance of a UFO signature during the October 30, 2007 observation does not seem to be related to specific disc-corona properties. 

 \begin{table}[t!]
\caption{SED fits (UV to 10\,keV) of the 2007\#4 {\sl XMM-Newton}-pn spectrum of ESO141-G55  with the baseline model {\sc relagn+reflkerrd}.} 
\centering
\begin{tabular}{@{}l c}
\hline\hline
Parameters &   2007\#4\\
\hline
$a$ & 0(f) \\
$\theta$ (degrees) & 43(f)\\
log\,$\dot{m}$ &  $-$1.01$^{+0.03}_{-0.02}$\\
$kT_{\rm hot}$ (keV)   & 136 (f)\\
$\tau_{\rm hot}$ & 0.47$^{+0.13}_{-0.07}$   \\
$\Gamma_{\rm hot}$ & 2.05$^{+0.05}_{-0.06}$\\
$R_{\rm hot}$ (R$_{\rm g}$) & 19$^{+2}_{-1}$\\
$kT_{\rm warm}$ (keV) & 0.40$^{+0.14}_{-0.10}$\\
$\Gamma_{\rm warm}$ & 2.82$^{+0.13}_{-0.09}$\\
$R_{\rm warm}$  (R$_{\rm g}$) & 51$^{+12}_{-8}$\\
log\,$\xi$  & 0.6(f)\\
$A_{\rm Fe}$ & 1.4(f)\\
$norm_{\rm reflkerrd}$ ($\times$10$^{-3}$) & 4.8$^{+0.7}_{-1.3}$\\
\hline
$F^{\rm unabs}_{\rm 0.001-100\,keV}$$^{(a)}$ (erg\,cm$^{-2}$\,s$^{-1}$) &   5.0$\times$10$^{-10}$ \\
$L^{\rm unabs}_{\rm 0.001-100\,keV}$$^{(a)}$ (erg\,s$^{-1}$)  &   1.7$\times$10$^{45}$ \\
$\chi^{2}$/d.o.f. & 518.3/440 \\
\hline    \hline
\end{tabular}
\label{tab:SED2007}
\flushleft
\small{{\bf Notes}. $^{(a)}$ Source flux and luminosity corrected from Galactic absorption and reddening. (f) means that the value was fixed to that found for the 2022 observation (Table~\ref{tab:SEDspin}).}
\end{table}

\end{document}